\documentclass[aps,pra,twocolumn,tightenlines,superscriptaddress,longbibliography]{revtex4-2}

\usepackage[utf8]{inputenc}
\usepackage{amsmath, amssymb, amsfonts}
\usepackage{graphicx}

\usepackage{physics}
\usepackage{epsfig}
\usepackage{color}

\allowdisplaybreaks

\usepackage{etoolbox}

\usepackage[colorlinks]{hyperref}
\hypersetup{%
        plainpages=true,
        breaklinks=true,
        hypertexnames=false,
        pageanchor=true,
        colorlinks=true,
        linkcolor={blue},
        citecolor={blue},
        urlcolor={blue},
        anchorcolor={black}
      }

\usepackage{mleftright} 

\newcommand{\figref}[1]{\mbox{Fig.~\ref{#1}}}

\newcommand{\secref}[1]{\mbox{Sec.~\ref{#1}}}

\newcommand{\appref}[1]{\mbox{Appendix~\ref{#1}}}

\newcommand{\figpanel}[2]{Fig.~\hyperref[#1]{\ref*{#1}(#2)}}
\newcommand{\figpanels}[3]{Fig.~\hyperref[#1]{\ref*{#1}(#2)--(#3)}}
\newcommand{\figpanelNoPrefix}[2]{\hyperref[#1]{\ref*{#1}(#2)}}

\usepackage{enumitem}   

\usepackage{cleveref}
\crefname{equation}{Eq.}{Eqs.}
\Crefname{equation}{Equation}{Equations}
\crefrangelabelformat{equation}{(#3#1#4--#5#2#6)}
\crefmultiformat{equation}{Eqs.~(#2#1#3}{, #2#1#3)}{#2#1#3}{#2#1#3}
\Crefmultiformat{equation}{Equations (#2#1#3}{, #2#1#3)}{#2#1#3}{#2#1#3}

\begin{document}

\title{Programmable Hong--Ou--Mandel interference in a giant-atom beam splitter}

\author{Ruolin Chai}
\affiliation{Center for Theoretical Physics \& School of Physics and Optoelectronic Engineering, Hainan University, Haikou 570228, China}

\author{Lei Du}
\email{lei.du@chalmers.se}
\affiliation{Department of Microtechnology and Nanoscience, Chalmers University of Technology, 412 96 Gothenburg, Sweden}

\author{Guoqing Cai}
\affiliation{Center for Theoretical Physics \& School of Physics and Optoelectronic Engineering, Hainan University, Haikou 570228, China}

\author{Alejandro Vivas-Via\~na}
\affiliation{Department of Microtechnology and Nanoscience, Chalmers University of Technology, 412 96 Gothenburg, Sweden}

\author{Anton Frisk Kockum}
\email{anton.frisk.kockum@chalmers.se}
\affiliation{Department of Microtechnology and Nanoscience, Chalmers University of Technology, 412 96 Gothenburg, Sweden}

\author{Yong Li}
\email{yongli@hainanu.edu.cn}
\affiliation{Center for Theoretical Physics \& School of Physics and Optoelectronic Engineering, Hainan University, Haikou 570228, China}


\begin{abstract}

The Hong--Ou--Mandel (HOM) effect is a hallmark of two-photon quantum interference, in which two indistinguishable photons impinging on a balanced beam splitter bunch into the same output port. Here, we show that a giant atom (GA), coupled to two waveguides through two coupling points each, can function as a programmable HOM interferometer, enabling continuous control over single-photon beam splitting and two-photon interference. This programmability arises from coupling-phase differences in the GA, which tune its self-interference and directionality, and thereby its scattering response. To characterize the two-photon interference, we analyze the scattering of two Gaussian single-photon wave packets injected through different waveguides and evaluate the bunching and antibunching (coincidence) probabilities of the resulting four output ports. At the operating point where the GA acts as an effective 50:50 beam splitter for single photons, we observe a pronounced HOM dip as the relative input delay between the two wave packets is varied. Away from this point, adjusting the coupling phases continuously tunes the two-photon output statistics between bunching into the same output port and antibunching across distinct output ports. In addition, we explore an application to quantum parameter estimation, showing that small deviations of coupling phases can be estimated from the two-photon output statistics, with the achievable sensitivity quantified by the classical Fisher information associated with a binary coincidence measurement. Our giant-atom beam splitter thus provides a programmable platform for two-photon interference in waveguide quantum electrodynamics, with potential applications in quantum information processing, quantum communication, and quantum sensing.

\end{abstract}

\date{\today}

\maketitle


\section{Introduction}

Two-photon processes provide a fundamental probe of nonclassical phenomena in quantum optics, directly revealing photon correlations and higher-order interference effects~\cite{Mandal1983PRA, HOM2_1987}. 
These processes also play a central role in photonic quantum technologies, where few-photon states and their correlations serve as key resources for encoding, transmitting, and processing quantum information~\cite{photonic_2009, divincenzo_physical_2000, Slussarenko2019APR}.
In particular, two-photon scattering probes nonlinear effects beyond the single-photon regime, including emitter saturation and correlated output statistics induced by light--matter interaction and effective optical nonlinearities~\cite{shen_strongly_2007, shi_two-photon_2011}.

A paradigmatic example of two-photon interference is the Hong--Ou--Mandel (HOM) effect~\cite{HOM_1987, HOM_Review}: when two indistinguishable photons arrive simultaneously at an ideal balanced beam splitter from different input ports, destructive interference suppresses coincidence events, causing both photons to emerge from the same output port. This interference produces the characteristic HOM dip in the coincidence probability as the relative arrival-time delay is tuned to zero. Beyond its fundamental connection to bosonic exchange symmetry, HOM interference has become a practical diagnostic of photon indistinguishability~\cite{EsmannSolidStateSinglePhoton2024} and a resource for calibrating linear-optical networks~\cite{calibratinglaing}, generating entanglement~\cite{HOMentanglement1993, HOMentanglement1998}, performing Bell-state measurements~\cite{BellMeasurement, LOQC_Review, Bouwmeester1997}, and implementing quantum interferometers~\cite{interferometer2024, metrologytiminglimits}.

In conventional HOM setups, the beam splitter is typically a passive linear-optical element whose splitting ratio is fixed or adjusted independently of the photon-scattering dynamics~\cite{HalfInsertedHOM, metasurfaceHOM, HOM2_1987}. 
For integrated few-photon devices, it is therefore desirable to replace such a passive element with programmable quantum scatterers, allowing photon routing, beam splitting, interference, and output correlations to be controlled \textit{in situ} within a single platform. 
Steps that have been taken in this direction include theoretical proposals based on a single two-level atom coupled to a waveguide~\cite{Roulet2016} or embedded in a cavity interrupting a waveguide~\cite{Oehri2015}, as well as superconducting-circuit implementations using several pumped Josephson junctions~\cite{Abdo2013} or a frequency-tunable transmon qubit~\cite{Huang2025}. However, these approaches either require comparatively complex architectures or offer limited control over the scattering response. 

\begin{figure}[t!]
\includegraphics[width=1.0\linewidth]{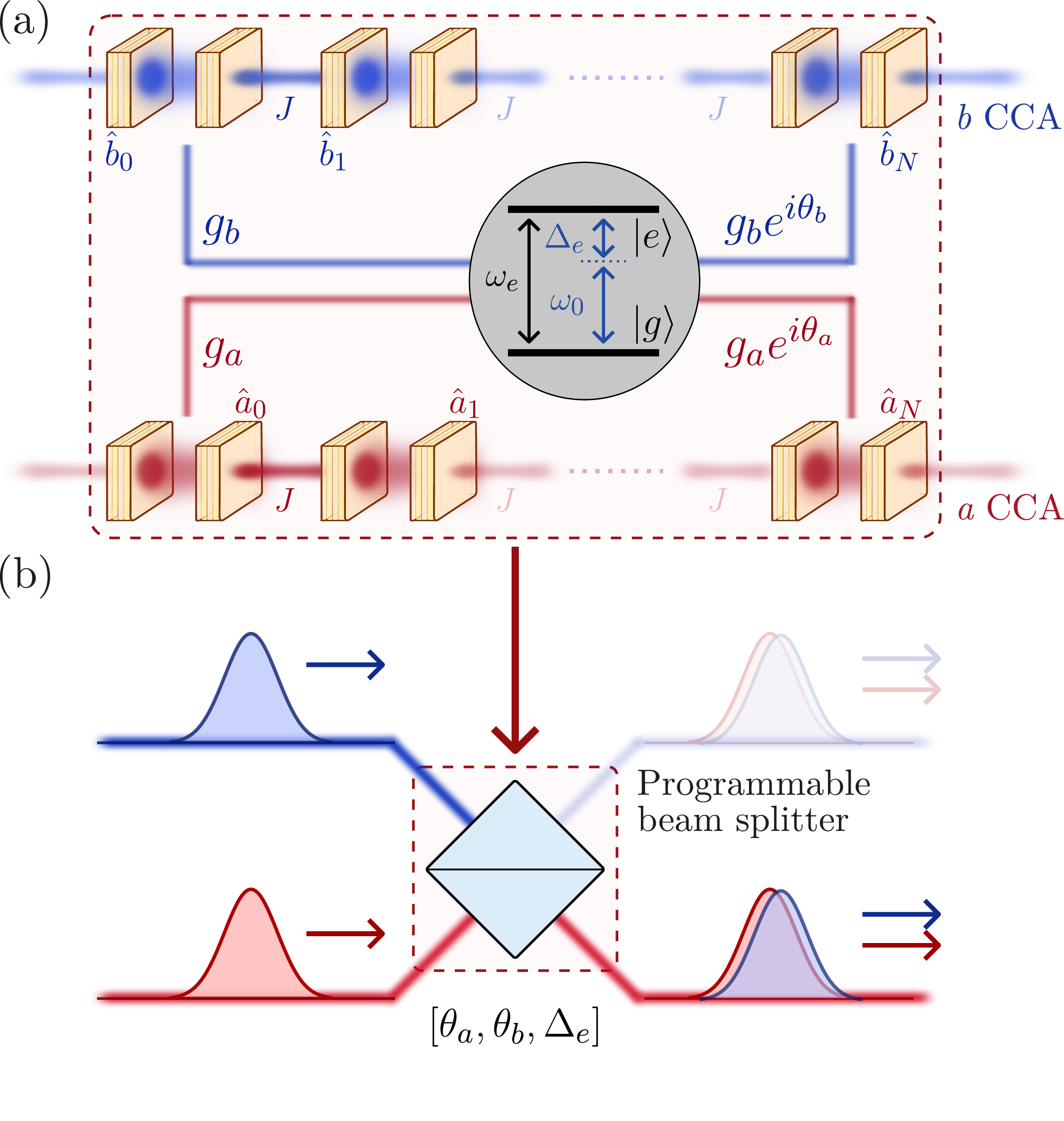}
\caption{Schematic of the system. 
(a) A two-level giant atom (GA) is coupled at two spatially separated points, $m=0$ and $m=N$, to two coupled-cavity arrays (CCAs), labeled $a$ and $b$, with coupling strengths $g_a$ and $g_b$, respectively. The couplings at $m=N$ acquire tunable phases $\theta_a$ and $\theta_b$. 
(b) Two single-photon wave packets incident through different CCAs scatter from the GA, which acts as a programmable beam splitter. Tuning the coupling phases $\theta_a$, $\theta_b$ and the GA-cavity detuning $\Delta_e$ controls the resulting four-port two-photon output.
\label{fig:Model}}
\end{figure}

Here, we propose and theoretically demonstrate an architecture for a programmable HOM interferometer based on giant atoms (GAs), i.e., artificial atoms coupled to a waveguide at multiple spatially separated points~\cite{GA.Theo, GA.summarize}. In GAs, the propagation phases accumulated between coupling points give rise to interference effects between different emission and reabsorption paths. These interference effects lead to characteristic phenomena such as tunable frequency-dependent relaxation rates and Lamb shifts~\cite{GA.Theo, GA.summarize}, allowing the \textit{in situ} enhancement or suppression of dissipation, as well as tunable interactions between GAs that can be protected from decay in both continuous waveguides and structured environments~\cite{GA.Theo2, GA.summarize, CarolloMechanismDecoherencefree2020,Soro_2023, Du2023DFIpra, Du2023SkinPRR, Ingelsten2024,LeonforteQuantumOptics2025}. These and other GA phenomena have been demonstrated in superconducting circuits~\cite{GA.Exp, Manenti2017, Andersson2019NP, meandering.Kannan, Vadiraj2021, JoshiResonanceFluorescence2023, Hu2024, Jouanny2025, Xiao2025, AlmanaklyDrivendissipativeEntanglement2026, Yang2026} and spin ensembles~\cite{Wang2022,You2026SA}. In these experiments, photon or phonon transport is governed by frequency-dependent light--matter interactions rather than by passive optical elements. Promising GA implementations based on cold-atom optical lattices~\cite{AGT2019GAprl}, Rydberg atoms~\cite{YTC2023PRR,YTC2024PRA}, and photonic resonator or waveguide arrays~\cite{LonghiGA,Lim2023PRA} have also been proposed.

To realize HOM interferometry with a GA, we exploit a recent insight: coupling-phase differences between coupling points can modify their interference effects and induce spatially asymmetric spontaneous emission, providing an experimentally accessible knob for realizing directional light--matter interactions~\cite{chen2022, WXqst2022, DLprl2022, JoshiResonanceFluorescence2023, DLqst2023, Suarez-ForeroChiralQuantum2025, Xu2026, Yang2026}. Such directional light--matter interfaces have enabled high-performance photon routing at the single- and few-photon level~\cite{Carlos2016PRA,chen2022, Ruolin_Routing}, frequency conversion~\cite{DLprrFC2021, DLprl2022}, quantum state transfer~\cite{WXqst2022, DLprl2025, WXdoublon2026, CarlosGApassive}, and two-photon quantum gates~\cite{QuantumGateAGT2025, QuantumGateAGT2026}. These advances position GAs as powerful building blocks for quantum interconnects and quantum information processing. However, exploiting this coupling-phase control to realize and tune HOM interference with GAs has so far remained largely unexplored.

In this work, we show that a GA can serve as a programmable HOM interferometer, with its single-photon beam-splitting response and two-photon output correlations continuously controlled by the coupling phases. We demonstrate this by analyzing single- and two-photon scattering in the setup shown in \figref{fig:Model}. The system consists of a single two-level emitter coupled to two waveguides in the form of coupled-cavity arrays (CCAs). The emitter is coupled to each CCA at two spatially separated sites, with tunable coupling-phase differences, $\theta_a$ and $\theta_b$. 

We first show that these phases tune the single-photon scattering response, enabling intra-waveguide transmission, inter-waveguide transfer, and reflection-free balanced beam splitting~\cite{chen2022, zhu2024}. Guided by these single-photon scattering properties of the system, we then simulate the scattering of two Gaussian single-photon wave packets, one incident through each waveguide, and analyze the asymptotic two-photon output. 
Since bidirectional propagation across the two waveguides defines a four-port scattering space, we characterize HOM interference through a generalized coincidence probability that includes all events in which the two photons exit through distinct ports. 

We find that, under the reflection-free balanced-splitting condition, the GA acts as an effective 50:50 beam splitter, yielding a clear HOM dip in the coincidence probability as the relative input delay is varied. 
Furthermore, tuning the coupling phases away from this operating point drives a continuous crossover between two-photon bunching and antibunching, demonstrating programmable control over the correlated output statistics. 
Building on this programmability, we use the framework of quantum parameter estimation~\cite{ParisQUANTUMESTIMATION2009, PetzINTRODUCTIONQUANTUM2011, Demkowicz-DobrzanskiQuantumLimits2015, PezzeQuantumMetrology2018, PolinoPhotonicQuantum2020} to show that the sensitivity of the HOM statistics to small phase deviations can be exploited for estimating the coupling phases, potentially enabling their \textit{in situ} calibration. 

We note that resolving the full two-photon output statistics benefits from photon-number-resolving detection, which remains challenging in superconducting circuits, the principal platforms for GA implementation~\cite{Gu2017, NarlaRobustConcurrent2016, OpremcakMeasurementSuperconducting2018, KonoQuantumNondemolition2018, BesseSingleShotQuantum2018, AlbertMicrowavePhotonNumber2024}. Nevertheless, microwave photon-counting efficiencies have improved substantially in recent years~\cite{May2025, Oppliger2026, Kulkarni2026}. Furthermore, correlation functions, as well as arbitrary moments and cumulants, can be extracted in superconducting circuits using quadrature measurements~\cite{Gu2017, EichlerCharacterizingQuantum2012, DaSilva2010, Bozyigit2011, Virally2016}. This method has been used to observe the HOM effect in superconducting circuits~\cite{LangCorrelationsIndistinguishability2013, Woolley2013} and should work well for observing it also in our proposed setup; further analysis is warranted for determining how this method would apply to the metrological scheme.

This article is organized as follows. In \secref{sec:model}, we introduce the model. Then, in \secref{sec:SinglePhotonScattering}, we analyze the single-photon scattering problem and identify the conditions for reflection-free balanced beam splitting. In \secref{sec:TwoPhotonScattering}, we extend the analysis to two-photon scattering and characterize HOM interference and the phase-dependent two-photon output statistics. In \secref{sec:metrology}, we develop a metrological protocol for coupling-phase estimation based on the HOM output statistics. Finally, we conclude in \secref{sec:conclusion}.


\section{Model}
\label{sec:model}

We model the GA as a two-level emitter with transition frequency $\omega_e$ and lowering operator $\hat \sigma=|g\rangle \langle e|$, where $|g\rangle$ and $|e\rangle$ denote its ground and excited states, respectively. The emitter is coupled to two infinite CCA waveguides, labeled $a$ and $b$ CCAs, as illustrated in \figpanel{fig:Model}{a}. We assume that all cavities in both waveguides have the same bare resonance frequency, $\omega_{a,m} = \omega_{b,m} \equiv \omega_0$. In each CCA, the GA couples to two cavities at sites $m=0$ and $m=N$, where $N \ge 1$ denotes the separation between the coupling points in lattice sites. Henceforth, we set $\hbar=1$. The bare Hamiltonian of the uncoupled GA and cavities reads
\begin{equation}
    \hat H_0=\omega_e \hat \sigma^{\dagger} \hat \sigma+\omega_0 \sum_{m=-\infty}^\infty \sum_{\alpha=a, b} \hat{\alpha}_m^{\dagger} \hat{\alpha}_m ,
\end{equation}
where $\hat{a}_m$ ($\hat{a}_m^\dag$) and $\hat{b}_m$ ($\hat{b}_m^\dag$) are the annihilation (creation) operators for cavity $m$ in the $a$ and $b$ lattices, respectively. Including the GA-CCA coupling and nearest-neighbor hopping within the CCAs, moving to a frame rotating at $\omega_0$, and applying the rotating-wave approximation, we obtain the system Hamiltonian
\begin{align}
    \hat H=\Delta_e \hat \sigma^\dagger\hat \sigma + \sum_{\alpha=a,b}\Big[&-J\sum_{m=-\infty}^\infty \hat{\alpha}_m^\dagger \hat{\alpha}_{m+1} \notag \\
    &+  g_\alpha\hat \sigma(\hat \alpha_0^\dagger+e^{i\theta_\alpha}\hat \alpha_N^\dagger) +\text{H.c.}\Big].
    \label{eq:H_int}
\end{align}
Here, the first term describes the GA energy relative to the CCA band center, with $\Delta_e\equiv\omega_e-\omega_0$ the GA--cavity detuning. The second term describes nearest-neighbor hopping in each CCA with uniform hopping rate $J$.  Finally, the last term describes the coupling between the GA and each CCA at the sites $m=0$ and $m=N$, with coupling strengths $g_a$ and $g_b$. 

The phase factors $e^{i\theta_a}$ and $e^{i\theta_b}$ in \cref{eq:H_int} encode tunable coupling-phase differences between the two coupling points in each CCA. Such coupling phases can be implemented experimentally using tunable couplers in superconducting circuits~\cite{JoshiResonanceFluorescence2023, Yang2026}.  In each CCA, the GA self-interference is governed by the total relative phase accumulated between its coupling points, which combines the tunable coupling phase $\theta_\alpha$ with the propagation phase $kN$, where $k$ is the wave vector of the propagating photon. This interference controls the frequency-dependent scattering response and can therefore be tuned \textit{in situ} through $\theta_\alpha$. Note that while individual coupling phases can be redistributed by local gauge transformations, the relative phases entering the interference between the two coupling paths have observable consequences for the GA scattering response.
 
For simplicity, we take the hopping rate $J$ and the GA--CCA coupling strengths $g_{\alpha}$ to be real and positive. Although we retain $g_a$ and $g_b$ as independent parameters in the general expressions below, all beam-splitter operating conditions and numerical results are evaluated for symmetric couplings, $g_a=g_b\equiv g$.


\section{Single-photon scattering}
\label{sec:SinglePhotonScattering}

We now explore the single-photon scattering in the setup shown in \figpanel{fig:Model}{a}. Since the system Hamiltonian in~\cref{eq:H_int} conserves the total excitation number, the single-photon dynamics are restricted to the single-excitation subspace. This subspace is spanned by $\{\ket{0,e},\, \hat{a}_m^\dagger\ket{0,g},\, \hat{b}_m^\dagger\ket{0,g}\}$,
where $\ket{0,g}$ and $\ket{0,e}$ denote the states with the emitter in its ground and excited states, respectively, with no photons in the CCAs. Here, we use the compact notation $|0\rangle\equiv|0_a,0_b\rangle$ for the joint vacuum state of the two CCAs. An eigenstate of the system in the single-excitation subspace has the general form
\begin{equation}
\ket{\Phi_1}\equiv\sum_m \mleft(A_m \hat{a}_m^\dagger+B_m \hat{b}_m^\dagger\mright)\ket{0,g}+C_e\ket{0,e}.
\label{Eq.SingleState}
\end{equation}
Here, $A_m$ ($B_m$) is the probability amplitude for a photon to occupy site $m$ of the $a$ CCA ($b$ CCA), while $C_e$ denotes the probability amplitude of the emitter being in its excited state.

Since each CCA supports both right- and left-propagating modes, the two-CCA system contains four asymptotic scattering ports. For a photon incident from the left side of the $a$ CCA, the outgoing waves moving to the right in the region $m>N$ define the transmission amplitudes $t_a$ and $t_b$ in the $a$ and $b$ CCAs, respectively, while the outgoing waves moving to the left in the region $m<0$ define the corresponding reflection amplitudes $r_a$ and $r_b$. Setting the incident amplitude to unity, we write the scattering ansatz~\cite{betheansatz2016, zhouRouting2013, ZhihaiBoundstate2020, dinc2019}
\begin{eqnarray}
A_{m} & = & \begin{cases}
e^{ikm}+r_{a}e^{-ikm}, & m\leqslant 0,\\[4pt]
\xi_{al}e^{-ikm}+\xi_{ar}e^{ikm}, & 0 < m < N,\\[4pt]
t_{a}e^{ikm}, & m\geqslant N,
\end{cases}
\label{eq:u_m}
\end{eqnarray}
and
\begin{eqnarray}
B_{m} & = & \begin{cases}
r_{b}e^{-ikm}, & m\leqslant 0,\\
\xi_{bl}e^{-ikm}+\xi_{br}e^{ikm}, & 0 < m < N,\\
t_{b}e^{ikm}, & m\geqslant N.
\end{cases}
\label{eq:v_m}
\end{eqnarray}
Here, the coefficients $\xi_{al}$ ($\xi_{bl}$) and $\xi_{ar}$ ($\xi_{br}$) denote the left- and right-propagating amplitudes, respectively, between the two coupling points in the $a$ CCA ($b$ CCA). The CCA dispersion relation, obtained from the hopping term in \cref{eq:H_int}, is given by $E_k=-2J\cos k$.

Substituting the scattering ansatz into the eigenvalue equation $\hat H\ket{\Phi_1}=E_k\ket{\Phi_1}$ and applying the continuity conditions at the coupling sites, we obtain the single-photon transmission and reflection amplitudes
\begin{subequations}\label{eq:transmission}
    \begin{align}
        t_a&=1+\frac{g_{a}}{2iJ\sin k}\mleft[1+e^{i\mleft(\theta_{a}-kN\mright)}\mright]C_{e},\label{eq:ta}\\
        t_b&=\frac{g_{b}}{2iJ\sin k}\mleft[1+e^{i\mleft(\theta_{b}-kN\mright)}\mright]C_{e},\label{eq:tb}\\
        r_a&=\frac{g_{a}}{2iJ\sin k}\mleft[1+e^{i\mleft(\theta_{a}+kN\mright)}\mright]C_{e},\label{eq:ra}\\
        r_b&=\frac{g_{b}}{2iJ\sin k}\mleft[1+e^{i\mleft(\theta_{b}+kN\mright)}\mright]C_{e},\label{eq:rb}
    \end{align}
\end{subequations}
and the GA excitation amplitude
\begin{equation}
C_{e}=\frac{g_{a}\mleft[1+e^{i\mleft(kN-\theta_{a}\mright)}\mright]}
{E_{k}-\Delta_{e}-D_{a}-D_{b}}.
\label{eq:Ce}
\end{equation}
Here, $D_{\alpha}$ denotes the on-shell retarded self-energy contribution from the $\alpha$ CCA~\cite{MarkovianBath2017, Soro_2023} (see details in Appendix~\ref{app:self_energy_two_lattices}):
\begin{equation}
D_{\alpha}=\frac{g_{\alpha}^{2}}{iJ\sin k}\mleft(1+e^{ikN}\cos\theta_{\alpha}\mright),\quad \alpha=a,b.
\label{eq:Dalpha}
\end{equation}
Note that the numerator of \cref{eq:Ce} only contains the coupling to the $a$ CCA because the incident photon is incident through this CCA, whereas both CCAs contribute to the self-energy in the denominator.

For a photon incident from the left of the GA in the $a$ CCA, the single-photon transmission probabilities into the forward (right-moving) output channels of the $a$ and $b$ CCAs are, respectively,
\begin{subequations}
    \label{eq:TaTb}
    \begin{align}
        T_{a\rightarrow a}&=\mleft|t_{a}\mright|^{2}, \label{eq:Ta}\\
        T_{a\rightarrow b}&=\mleft|t_{b}\mright|^{2}.\label{eq:Tb}
    \end{align}
\end{subequations}
These transmission probabilities allow us to identify the operating conditions under which the GA acts as an effective 50:50 beam splitter for single photons, requiring both complete suppression of reflection and equal splitting of the incident photon between the two forward output channels. 

In the absence of intrinsic dissipation, as assumed throughout this work, complete suppression of reflection implies $T_{a\rightarrow a}+T_{a\rightarrow b}=1$. For $\theta_a=\theta_b\equiv\theta$, this condition is satisfied when
\begin{equation}
    \theta \pm kN = (2q + 1)\pi, \quad q\in \mathbb{Z}.
    \label{eq:phase-condition}
\end{equation}
Further imposing the condition of balanced splitting, $T_{a\rightarrow a}=T_{a\rightarrow b}=1/2$, selects the branch $\theta + kN = (2q + 1)\pi$. Combining this condition with \cref{eq:ta,eq:tb} along with \crefrange{eq:Ce}{eq:TaTb}, we obtain the GA--cavity detuning
\begin{equation}
\Delta_e=-2J\cos k+\frac{g^2}{J\sin k}\mleft[\sin(2kN)\pm 2\sin^2(kN)\mright],
\label{eq:ConditionDeltae}
\end{equation}
where we have assumed $g_a=g_b\equiv g$. These balanced-splitting solutions require $0<k<\pi$ and $\sin(kN)\neq0$.  When $\sin(kN)=0$, the effective forward coupling vanishes,  and the apparent solution is singular rather than corresponding to a physical 50:50 operating point. For a photon at the band center, $k=\pi/2$, and a GA coupled to adjacent cavities, $N=1$, \cref{eq:ConditionDeltae} reduces to $\Delta_e=\pm2g^2/J$.

\begin{figure}[t!]
\includegraphics[width=1\linewidth]{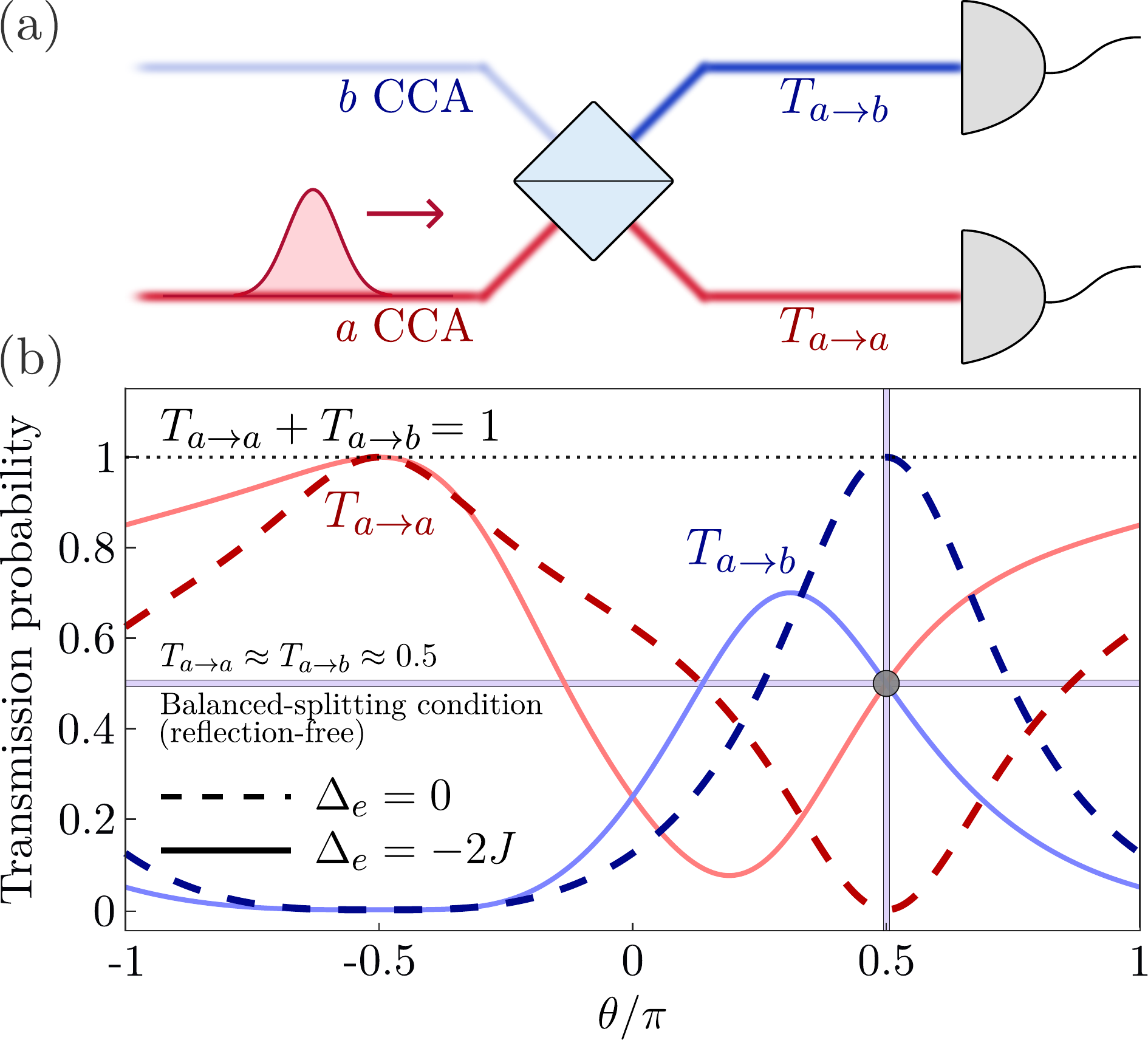}
\caption{Phase-controlled single-photon transmission. 
(a) Schematic of single-photon scattering for a photon incident through the $a$ CCA. The transmitted photon can exit through either the $a$ or $b$ CCA, with transmission probabilities $T_{a\to a}$ and $T_{a\to b}$, respectively.
(b) Transmission probabilities $T_{a\to a}$ (red) and $T_{a\to b}$ (blue) as a function of the common coupling phase $\theta_a=\theta_b\equiv\theta$, for detunings $\Delta_e=0$ (dashed lines) and $\Delta_e=-2J$ (solid lines). The dotted horizontal line indicates the reflection-free condition $T_{a\to a}+T_{a\to b}=1$. 
The horizontal and vertical shaded regions indicate $T_{a\to a}\approx T_{a\to b}\approx 0.5$ (under the reflection-free condition) and $\theta\approx\pi/2$, respectively, while the filled circle marks the balanced-splitting point at $T_{a\to a}=T_{a\to b}=0.5$ and $\theta=\pi/2$.
Parameters: $k=\pi/2$, $g_a=g_b=J$, and $N=1$. }
\label{fig:Tab}
\end{figure}

In \figref{fig:Tab}, we illustrate the dependence of the single-photon transmission probabilities on the coupling phase $\theta$ and GA-cavity detuning $\Delta_e$. For the resonant case, $\Delta_e=0$ (dashed curves), the reflection-free condition is reached at $\theta=\pi/2$, where the incident photon is completely transferred from the $a$ to the $b$ CCA, with $T_{a\to b}=1$ (blue dashed) and $T_{a\to a}=0$ (red dashed). In contrast, at $\theta=-\pi/2$, the photon remains in the $a$ CCA, with $T_{a\to a}=1$ and $T_{a\to b}=0$. For $\Delta_e=-2J$ (solid curves), the detuning modifies the distribution between the two transmission channels, allowing the probabilities to become equal. In particular, at $\theta=\pi/2$, we obtain $T_{a\to a}=T_{a\to b}=1/2$, realizing the reflection-free 50:50 beam-splitter condition predicted by~\cref{eq:phase-condition,eq:ConditionDeltae}. These results illustrate the complementary roles of the coupling phase and GA–cavity detuning in controlling the reflection and distribution of the scattered photon, with both conditions required to realize reflection-free balanced splitting.

The balanced-splitting detuning admits a simple effective interpretation. Once $\theta$ is chosen according to \cref{eq:phase-condition} to suppress  reflection, the detuning $\Delta_e$ determines how the transmitted photon is distributed between the two CCAs. This reduced problem can be mapped onto a single-photon scattering problem of an emitter in a single waveguide~\cite{Shen2005}: the transmission into the $a$ CCA corresponds to the transmission in the effective single waveguide while transmission into the $b$ CCA corresponds to the reflection. For $k=\pi/2$, the photon energy relative to the CCA band center is $E_k=0$; thus $\Delta_e=0$ corresponds to resonance between the photon and the emitter. A resonant photon ($\Delta_e=0$) is then completely transferred from the $a$ to the $b$ CCA, as shown by the blue dashed curve in \figpanel{fig:Tab}{b}. Balanced splitting is obtained by detuning the emitter from resonance by half the effective linewidth.  Since the total decay rate into the effective waveguide is  $4g^2/J$, this yields the 50:50 beam-splitter condition $\Delta_e=\pm2g^2/J$, in agreement with \cref{eq:ConditionDeltae} (see details in \appref{app:self_energy_two_lattices}).

Single-photon routing among different waveguides has been widely explored using atom-based and hybrid systems~\cite{Carlos2016PRA,Gu2017, WaveguideQED_Review, Hoi2011, zhouRouting2013, Zhoulan, Lu2015OE, CHYan2018PRA, Ahumada2019PRA, GAYan2020QST, Xu2021PR, Singh2023PRR, XZhang2024routing}. In particular, interference between spatially separated coupling points provides GA-based schemes with enhanced tunability, allowing photon transmission and frequency conversion to be controlled through the coupling geometry and phases~\cite{chen2022,Ruolin_Routing,Gong2024}. Our results further confirm that the same interference can configure the GA as a programmable beam splitter. We now turn to the two-photon regime and explore how this tunability controls the routing and output correlations of two incident photons.


\section{Two-photon scattering}
\label{sec:TwoPhotonScattering}

Having established the single-photon scattering framework and the conditions for reflection-free single-photon beam splitting, we now turn to the two-photon scattering problem and explore HOM interference. 


\subsection{Two-photon scattering state}

The two-photon dynamics are confined to the two-excitation subspace, in which the general time-dependent scattering state can be expanded as
\begin{widetext}
\begin{equation}
\ket{\Phi_2}\equiv\mleft[\sum_{m=-\infty}^\infty \sum_{n=-\infty}^m \mleft(C_{m,n}^{aa}\frac{\hat a_m^\dagger \hat a_n^\dagger}{\sqrt{1+\delta_{m,n}}}+C_{m,n}^{bb}\frac{\hat b_m^\dagger \hat b_n^\dagger}{\sqrt{1+\delta_{m,n}}}\mright)+\sum_{m,n=-\infty}^\infty C_{m,n}^{ab}\hat a_m^\dagger \hat b_n^\dagger\mright]\ket{0,g} + \sum_m\mleft(C_m^{ea}\hat a_m^\dagger+C_m^{eb}\hat b_m^\dagger \mright)\ket{0,e}.
\label{eq:TwoPhotonState}
\end{equation}
\end{widetext}
Here, the terms with coefficients $C_{m,n}^{aa}$ and $C_{m,n}^{bb}$ describe the sectors in which there are two photons in the same CCA. 
Specifically, $C_{m,n}^{aa}$ ($C_{m,n}^{bb}$) is the amplitude for finding two photons in the $a$ CCA ($b$ CCA) at sites $m$ and $n$. Bosonic exchange symmetry allows these sectors to be restricted to $m\geq n$, avoiding double counting, while the factor $1/\sqrt{1+\delta_{m,n}}$ ensures proper normalization when both photons occupy the same site.
Terms with coefficient $C_{m,n}^{ab}$ describe the sector in which there is one photon in each CCA, giving the amplitude for finding one photon at site $m$ in the $a$ CCA and another at site $n$ in the $b$ CCA. Since the two indices refer to different CCAs, no ordering condition between $m$ and $n$ is imposed in this sector.
Finally, the terms with $C_m^{ea}$ and $C_m^{eb}$ describe the sectors in which the GA is excited and the remaining photon resides at site $m$ in the $a$ and $b$ CCAs, respectively.

Unlike in the single-photon case, the two-photon scattering solution cannot be readily obtained from a closed set of algebraic matching equations. Instead, we formulate the coupled equations of motion in the full two-excitation subspace and impose properly symmetrized initial conditions, as detailed in \appref{app:eom}. We then solve these equations numerically to explore phase-dependent two-photon routing and HOM interference.

Although our model allows the coupling phases $\theta_a$ and $\theta_b$ to be tuned independently, we first focus on the synchronized configuration, $\theta_a=\theta_b\equiv\theta$. This choice reduces the parameter space to a single control variable, thereby simplifying the analysis of the phase-dependent two-photon response. Experimentally, such a configuration can be realized by driving both tunable couplers with a common external flux~\cite{JoshiResonanceFluorescence2023, Yang2026}.

\begin{figure*}[t!]
\centering 
\includegraphics[width=1\linewidth]{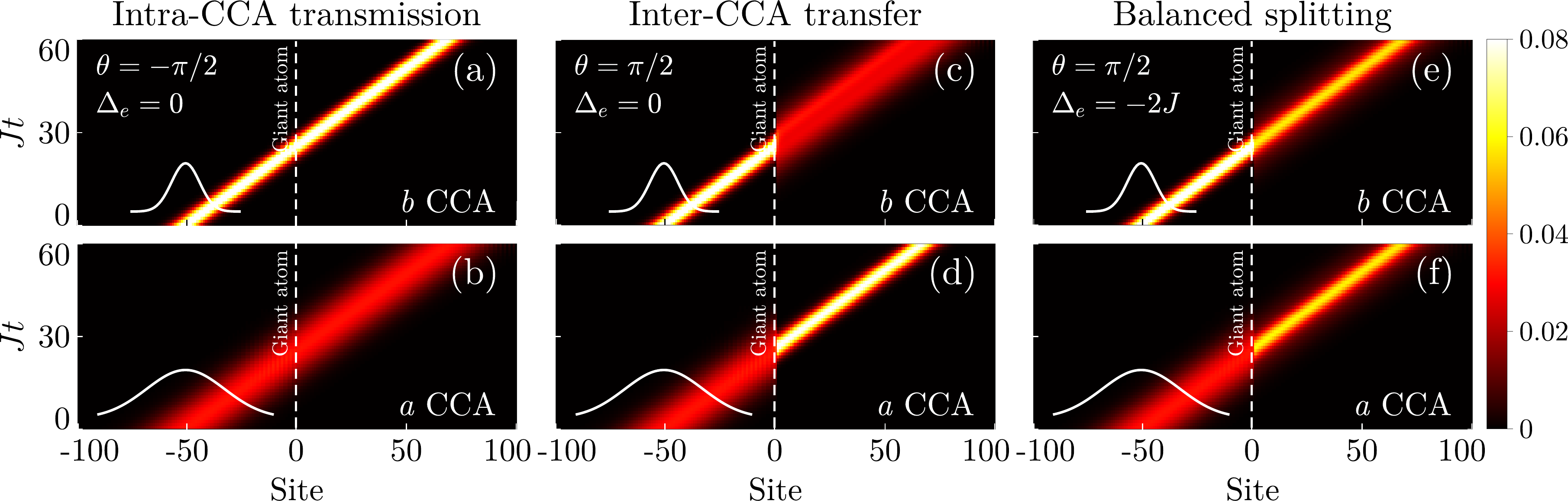}
\caption{Time evolution of the local photon occupations in the two CCAs for two incident Gaussian single-photon wave packets. Different widths, $w_a=25$ and $w_b=9$, are used to visually distinguish the two wave packets. The upper and lower panels show the occupations in the $b$ and $a$ CCAs, respectively.
(a, b) Intra-CCA transmission for $\theta=-\pi/2$ and $\Delta_e=0$. 
(c, d) Inter-CCA transfer for $\theta=\pi/2$ and $\Delta_e=0$. 
(e, f) Balanced splitting for $\theta=\pi/2$ and $\Delta_e=-2J$, corresponding to the single-photon 50:50 beam-splitter condition. 
The white curves indicate the initial spatial profiles of the incident wave packets, while the vertical dashed line marks the location of the giant atom. Parameters: $x_a=x_b=-50$, $N=1$, $g_a=g_b=J$, and $k_a=k_b=\pi/2$.}
\label{fig:EVO}
\end{figure*}


\subsection{Wave-packet dynamics}
\label{sec:DynamicalEvo}

We explore the two-photon scattering dynamics by injecting two Gaussian single-photon wave packets from the left, one through each CCA. At the initial time $t=0$, the wave packets are created by the single-photon creation operators
\begin{subequations}
    \begin{align}
        \hat\phi^{(1)\dagger}(0)&=\sum_{m=-\infty}^\infty A_m^{(1)}(0)\hat a_m^\dagger , \\
        \hat\phi^{(2)\dagger}(0)&=\sum_{n=-\infty}^\infty B_n^{(2)}(0)\hat b_n^\dagger,
    \end{align}
\end{subequations}
with the GA initially prepared in its ground state.
The corresponding Gaussian envelopes are
\begin{subequations}
    \begin{align}
        A_m^{(1)}(0)&=\mathcal{N}_a \exp\mleft[-(m-x_a)^2/(4w_a^2)+ik_a m\mright], \\
        B_n^{(2)}(0)&=\mathcal{N}_b \exp\mleft[-(n-x_b)^2/(4w_b^2)+ik_b n\mright].
    \end{align}
\end{subequations}
Here, $x_a$ and $x_b$ denote the initial wave-packet centers, chosen sufficiently far to the left of the GA coupling region, while $w_a$ and $w_b$ determine their spatial widths, and $k_a$ and $k_b$ their carrier wave vectors in the two CCAs. The normalization constants $\mathcal{N}_a$ and $\mathcal{N}_b$ ensure that each wave packet contains one photon.
The resulting incident two-photon state can be written as
\begin{equation}
    \ket{\Phi_2(0)}=\sum_{m, n}  A_m^{(1)}(0) B_n^{(2)}(0) \hat{a}_m^{\dagger} \hat{b}_n^{\dagger}\ket{0,g}.
\end{equation}

To characterize the scattering dynamics, we study the local photon occupations $P_{a,m}$ and $P_{b,m}$ in the two CCAs, together with the excited-state population $P_e$ of the GA. In terms of the two-excitation amplitudes, these quantities are given by
\begin{subequations}
\begin{align}
P_{a,m}(t)=&\bra{\Phi_2(t)}\hat{a}_m^{\dagger} \hat{a}_m\ket{\Phi_2(t)}\notag\\
=&\sum_{n=-\infty}^{m}\mleft|C_{m,n}^{aa}(t)\mright|^2 +\sum_{n=m}^{\infty}\mleft|C_{n,m}^{aa}(t)\mright|^2  \notag\\
&+\sum_{n=-\infty}^{\infty}\mleft|C_{m,n}^{ab}(t)\mright|^2 +\mleft|C_m^{ea}(t)\mright|^2,\\
P_{b,m}(t)=&\bra{\Phi_2(t)}\hat{b}_m^{\dagger} \hat{b}_m\ket{\Phi_2(t)}\notag\\
=&\sum_{n=-\infty}^{m}\mleft|C_{m,n}^{bb}(t)\mright|^2 +\sum_{n=m}^{\infty}\mleft|C_{n,m}^{bb}(t)\mright|^2  \notag\\
&+\sum_{n=-\infty}^{\infty}\mleft|C_{n,m}^{ab}(t)\mright|^2
+\mleft|C_m^{eb}(t)\mright|^2,\\
P_e(t)=&\bra{\Phi_2(t)}\hat{\sigma}^{\dagger} \hat{\sigma}\ket{\Phi_2(t)}\notag\\
=&\sum_{m=-\infty}^{\infty}\mleft|C_m^{ea}(t)\mright|^2
+\sum_{m=-\infty}^{\infty}\mleft|C_m^{eb}(t)\mright|^2 .
\end{align}
\end{subequations}

In \figref{fig:EVO}, we show the time evolution of the local photon occupations $P_{a,m}(t)$ and $P_{b,m}(t)$ for three representative parameter regimes identified from the single-photon scattering results in \figref{fig:Tab}. For visual clarity, we choose different wave-packet widths, $w_a=25$ and $w_b=9$, for the photons injected through the $a$ and $b$ CCAs, respectively.

For $\theta=-\pi/2$ and $\Delta_e=0$, Figs.~\figpanelNoPrefix{fig:EVO}{a,b} show that both wave packets propagate predominantly along the CCAs through which they were injected. This behavior is consistent with the single-photon scattering result presented in \figref{fig:Tab}, where these parameters correspond to perfect intra-CCA transmission.
By contrast, for $\theta=\pi/2$ and $\Delta_e=0$, Figs.~\figpanelNoPrefix{fig:EVO}{c,d} show that the two wave packets are transferred predominantly between the CCAs: the wave packet incident through the $a$ CCA ($b$ CCA) emerges mainly through the $b$ CCA ($a$ CCA). This inter-CCA transfer is again consistent with the corresponding single-photon scattering behavior in \figref{fig:Tab}.
Finally, Figs.~\figpanelNoPrefix{fig:EVO}{e,f} show the balanced-splitting regime, with $\theta=\pi/2$ and $\Delta_e=-2g^2/J=-2J$ for $g_a=g_b=J$. For these parameters, the single-photon results in \figref{fig:Tab} give $T_{a\to a}=T_{a\to b}=1/2$. Consistently, the local occupations show that each incident wave packet is approximately equally redistributed between the two CCAs after scattering. 

The results here thus provide a direct wave-packet demonstration of the effective beam-splitting action of the GA scatterer for two photons. However, these single-particle occupations do not reveal the joint two-photon output statistics and therefore cannot by themselves characterize HOM interference. Next, we therefore analyze the two-photon correlations and coincidence probability to demonstrate the HOM effect.


\subsection{Programmable Hong--Ou--Mandel interference}
\label{sec:HOM}

As shown in Secs.~\ref{sec:SinglePhotonScattering} and \ref{sec:DynamicalEvo}, for $N=1$ and $k=\pi/2$, the GA realizes reflection-free balanced single-photon splitting when $\theta_a=\theta_b\equiv\theta=\pi/2$ and $\Delta_e=\pm2g^2/J$.  We now explore the two-photon interference arising at this operating point and show that the GA realizes HOM interference, whose output statistics can be continuously controlled through the coupling phase $\theta$. The extension to larger coupling-point separations, $N>1$, is discussed in Appendix~\ref{app:alternative_separation}.

In conventional HOM interference, two indistinguishable photons incident through different input ports of a balanced beam splitter interfere destructively in the coincidence channels, suppressing events in which the photons exit through distinct output ports~\cite{HOM_1987,HOM_Review}. For perfectly indistinguishable photons---identical in polarization, spectrum, temporal profile, and arrival time--- an ideal lossless beam splitter therefore produces complete bunching into the same output port.

\begin{figure}[b!]
\centering 
\includegraphics[width=1\linewidth]{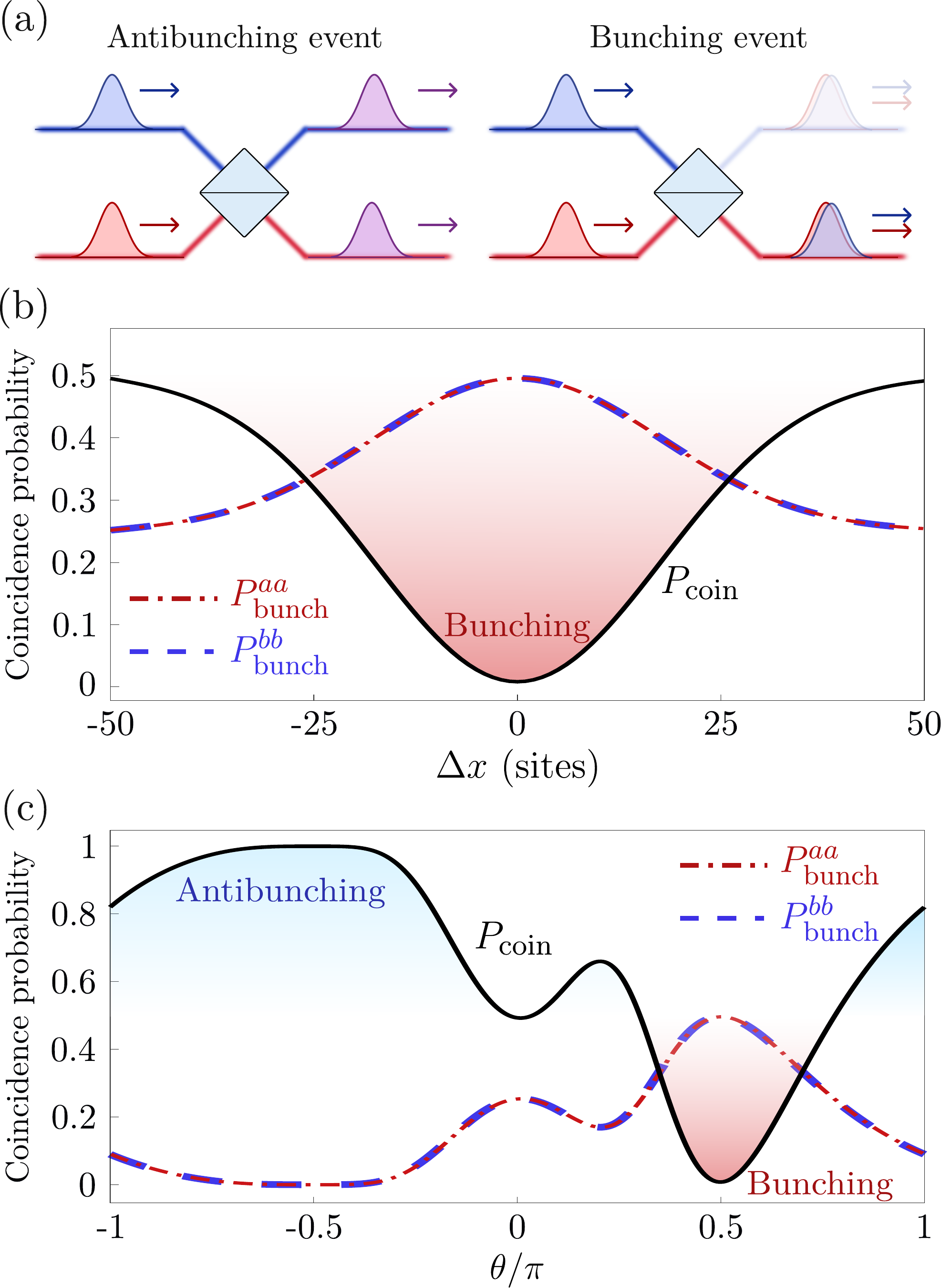}
\caption{Programmable HOM interference.
(a) Representative bunching and antibunching events in the four-port output geometry. 
(b) Generalized coincidence probability $P_{\rm coin}$ and bunching probabilities $P_{\rm bunch}^{aa}$ and $P_{\rm bunch}^{bb}$ as functions of the initial relative displacement $\Delta x=x_a-x_b$ between two identical Gaussian wave packets, one incident through each CCA, at the balanced-splitting point $\theta=\pi/2$ and $\Delta_e=-2J$.
(c) Same probabilities as functions of the coupling phase $\theta$ for fully overlapping inputs ($x_a=x_b=-50$). 
The shaded regions indicate bunching- and antibunching-dominated regimes.
Parameters: $N=1$, $w_a=w_b=25$, $g_a=g_b=J$, and $k_a=k_b=\pi/2$.
\label{fig:HOM-dip-theta}}
\end{figure}

In the GA setup considered here [see \figpanel{fig:Model}{a}], the effective beam-splitting amplitudes can be tuned through the coupling phases, providing an additional control knob beyond the indistinguishability of the incident photons. Moreover, because each CCA supports both left- and right-propagating modes, the scattering geometry contains four output ports. We therefore define a coincidence (antibunching) event as any outcome in which the two photons exit through distinct output ports, irrespective of whether these ports belong to the same or different CCAs, as illustrated schematically in \figpanel{fig:HOM-dip-theta}{a}.

To quantify HOM interference in this four-port geometry, we define the projector onto the coincidence (antibunching) subspace as
\begin{equation}
    \hat P_{\rm coin}\equiv\hat P^{ab}+\hat P^{aa}_{RL}+\hat P^{bb}_{RL}.
\end{equation}
The first contribution, 
\begin{equation}
    \hat P^{ab}\equiv\sum_{m,n=-\infty}^\infty  \hat{a}_m^{\dagger} \hat{b}_n^{\dagger}|0, g\rangle\langle0, g| \hat{a}_m \hat{b}_n,
    \label{eq:Pab}
\end{equation}
projects onto all states containing one photon in each CCA, which necessarily correspond to distinct output ports after scattering. A representative case, in which both photons exit through the right ends of their respective CCAs, is illustrated in the left panel of~\figpanel{fig:HOM-dip-theta}{a}.
The remaining two terms,
\begin{subequations}
    \begin{align}
        \hat P^{aa}_{RL}&\equiv\sum_{m=-\infty}^0 \sum_{n=N}^\infty\hat a_m^\dagger \hat a_n^\dagger\ket{0,g}\bra{0,g}\hat a_n \hat a_m, 
        \label{eq:PaaRL}
        \\
        \hat P^{bb}_{RL}&\equiv\sum_{m=-\infty}^0 \sum_{n=N}^\infty\hat b_m^\dagger \hat b_n^\dagger\ket{0,g}\bra{0,g}\hat b_n \hat b_m,
        \label{eq:PbbRL}
    \end{align}
\end{subequations}
project onto states in which both photons occupy the same CCA but propagate in opposite directions, and therefore also exit through distinct ports. Note that the summations over $m$ and $n$ cover different parts of the CCAs to enforce this condition.
We denote the asymptotic two-photon output state by $\ket{\Phi_{\rm out}}\equiv \ket{\Phi_2(t_f)}$, where the final time $t_f$ is chosen such that both wave packets have propagated beyond the GA coupling region and the GA excitation is negligible. The generalized coincidence probability is then
\begin{equation}
P_{\rm coin}
=
\bra{\Phi_{\rm out}}\hat P_{\rm coin}\ket{\Phi_{\rm out}} .
\label{eq.Pcoin}
\end{equation}
We use this generalized coincidence probability to examine the HOM interference under the single-photon balanced-scattering condition identified in \secref{sec:SinglePhotonScattering}.

Figure~\figpanelNoPrefix{fig:HOM-dip-theta}{b} shows the generalized coincidence probability as a function of the initial relative displacement $\Delta x\equiv x_a-x_b$ for two otherwise identical Gaussian wave packets at the balanced-splitting point $\theta=\pi/2$ and $\Delta_e=-2J$. Since the photons have identical carrier wave vectors and wave-packet widths, varying $\Delta x$ directly controls their relative arrival time and hence their temporal overlap at the GA. At $\Delta x=0$, the wave packets arrive simultaneously and $P_{\rm coin}$ exhibits a pronounced HOM dip, approaching zero for perfectly overlapping inputs.
The projectors onto states with both photons in the same output port are
\begin{subequations}
\begin{align}
\hat P^{\alpha\alpha}_{LL}& \equiv \sum_{m=-\infty}^{0}\sum_{n=-\infty}^{m}\frac{\hat\alpha_m^\dagger\hat\alpha_n^\dagger\ket{0,g}\bra{0,g}\hat\alpha_n\hat\alpha_m}{1+\delta_{m,n}},
\\
\hat P^{\alpha\alpha}_{RR}& \equiv \sum_{m=N}^{\infty}\sum_{n=N}^{m}\frac{\hat\alpha_m^\dagger\hat\alpha_n^\dagger\ket{0,g}\bra{0,g}\hat\alpha_n\hat\alpha_m}{1+\delta_{m,n}}.
\end{align}
\end{subequations}
The corresponding total bunching probability in the $\alpha$ CCA is
\begin{equation}
P^{\alpha\alpha}_{\rm bunch} \equiv \bra{\Phi_{\rm out}} \hat P^{\alpha\alpha}_{LL}+\hat P^{\alpha\alpha}_{RR} \ket{\Phi_{\rm out}},
\end{equation}
Here, $P^{\alpha\alpha}_{LL}$ and $P^{\alpha\alpha}_{RR}$ denote the probabilities that both photons exit through the left and right ports of CCA $\alpha$, respectively. The probability $P_{RR}^{\alpha\alpha}$ is illustrated in the right panel of \figpanel{fig:HOM-dip-theta}{a}.
At zero input delay, $P^{aa}_{\rm bunch}$ and $P^{bb}_{\rm bunch}$ approach $1/2$ in~\figpanel{fig:HOM-dip-theta}{b}. 

We next exploit the coupling phase $\theta$ as a control parameter for the two-photon interference. Figure~\figpanelNoPrefix{fig:HOM-dip-theta}{c} shows $P_{\rm coin}$ together with the bunching probabilities $P^{aa}_{\rm bunch}$ and $P^{bb}_{\rm bunch}$ as functions of $\theta$ for fully overlapping input wave packets ($\Delta x=0$). As $\theta$ is varied, $P_{\rm coin}$ can be tuned over nearly its full range, from almost unity to close to zero, while the bunching probabilities display the complementary behavior.  In particular, around $\theta=\pi/2$, the coincidence probability is strongly suppressed and the bunching probabilities approach $1/2$, recovering the HOM operating point identified in~\figpanel{fig:HOM-dip-theta}{b}. Away from this point, the output can instead become predominantly antibunched, with the photons preferentially exiting through distinct output ports.  Thus, tuning the coupling phase continuously controls the two-photon output from bunching to antibunching, demonstrating the programmable character of the GA HOM interferometer.


\section{Phase estimation via Hong--Ou--Mandel statistics}
\label{sec:metrology}

The self-interference of the GA is governed by the phases accumulated between its coupling points [see \figpanel{fig:Model}{a}], which combine the propagation phase ($k N$) with the externally tunable coupling phase $\theta_\alpha$. As discussed in \secref{sec:model}, the latter can be engineered using tunable couplers, such as SQUIDs controlled by an external magnetic flux~\cite{JoshiResonanceFluorescence2023,Cao2024,Peropadre2013,Ma2025}. In practice, however, accurately setting and stabilizing these phases can be challenging. Small deviations arising from environmental fluctuations or control drifts can alter the interference condition and degrade phase-sensitive protocols, including entanglement generation~\cite{AlmanaklyDrivendissipativeEntanglement2026}, quantum state transfer~\cite{WXqst2022,DLprl2025,WXdoublon2026,CarlosGApassive}, and photon routing~\cite{Carlos2016PRA,chen2022,Zhoulan,Ruolin_Routing}. This sensitivity motivates scattering-based strategies for estimating coupling-phase deviations, thereby enabling precise calibration of the coupling phases.

We consider a coupling-phase estimation protocol based on the HOM measurement. Specifically, we assume that the coupling phase $\theta_a$ is well characterized at the reference value $\theta_a=\pi/2$, while $\theta_b$ may deviate from its nominal value, treating it as the unknown parameter. The phase $\theta_b$ is encoded in the asymptotic two-photon state through the phase-dependent self-interference of the GA.
Rather than probing the tunable coupler directly, we infer $\theta_b$ from the scattered-photon statistics, since variations in this parameter are converted by the HOM interference into measurable changes in $P_{\rm coin}$.

We quantify the phase-estimation sensitivity using standard tools for quantum parameter estimation~\cite{ParisQUANTUMESTIMATION2009, PetzINTRODUCTIONQUANTUM2011, Demkowicz-DobrzanskiQuantumLimits2015, PezzeQuantumMetrology2018, PolinoPhotonicQuantum2020}.
The measurement process is described by a positive-operator-valued measure (POVM) $\hat\Lambda=\{\hat\Lambda_x\}$, where $x$ labels the possible measurement outcomes, with conditional probabilities
\begin{equation}
P(x|\theta_b)=\langle \Phi_{\rm out}(\theta_b)|\hat\Lambda_x|\Phi_{\rm out}(\theta_b)\rangle.
\label{Eq.POVM.prob}
\end{equation}
Here, $\ket{\Phi_{\rm out}(\theta_b)}$ denotes the normalized asymptotic two-photon output state,  treated as a function of the unknown coupling phase $\theta_b$. For the binary HOM measurement considered here, the POVM has two outcomes: antibunching and bunching, represented within the two-photon output subspace by
\begin{equation}
\hat{\Lambda}_{\rm HOM}=\mleft\{\hat P_{\rm coin},\hat{\mathbb{I}}-\hat P_{\rm coin}\mright\}.
\end{equation}
The projector $\hat P_{\rm coin}$ corresponds to the antibunching outcome, while its complement $\hat{\mathbb{I}}-\hat P_{\rm coin}$ represents bunching, where $\hat{\mathbb{I}}$ denotes the identity operator. 

To quantify the phase sensitivity of our measurement scheme, we compute the classical Fisher information (CFI)~\cite{ParisQUANTUMESTIMATION2009, PetzINTRODUCTIONQUANTUM2011, Demkowicz-DobrzanskiQuantumLimits2015, PezzeQuantumMetrology2018, PolinoPhotonicQuantum2020},
\begin{equation}
F_{\theta_b}=\sum_x\frac{1}{P(x|\theta_b)}\mleft(\frac{\partial P(x|\theta_b)}{\partial \theta_b}\mright)^2,
\label{eq:Fi}
\end{equation}
where the sum runs over the possible measurement outcomes. For the binary HOM measurement, this expression reduces to
\begin{equation}
F_{\theta_b}^{\rm HOM}=
\frac{\mleft[\partial_{\theta_b}P_{\rm coin}(\theta_b)\mright]^2}
{P_{\rm coin}(\theta_b)\mleft[1-P_{\rm coin}(\theta_b)\mright]} .
\label{eq:Fi_HOM}
\end{equation}
%

\begin{figure}[b!]
\centering
\includegraphics[width=\linewidth]{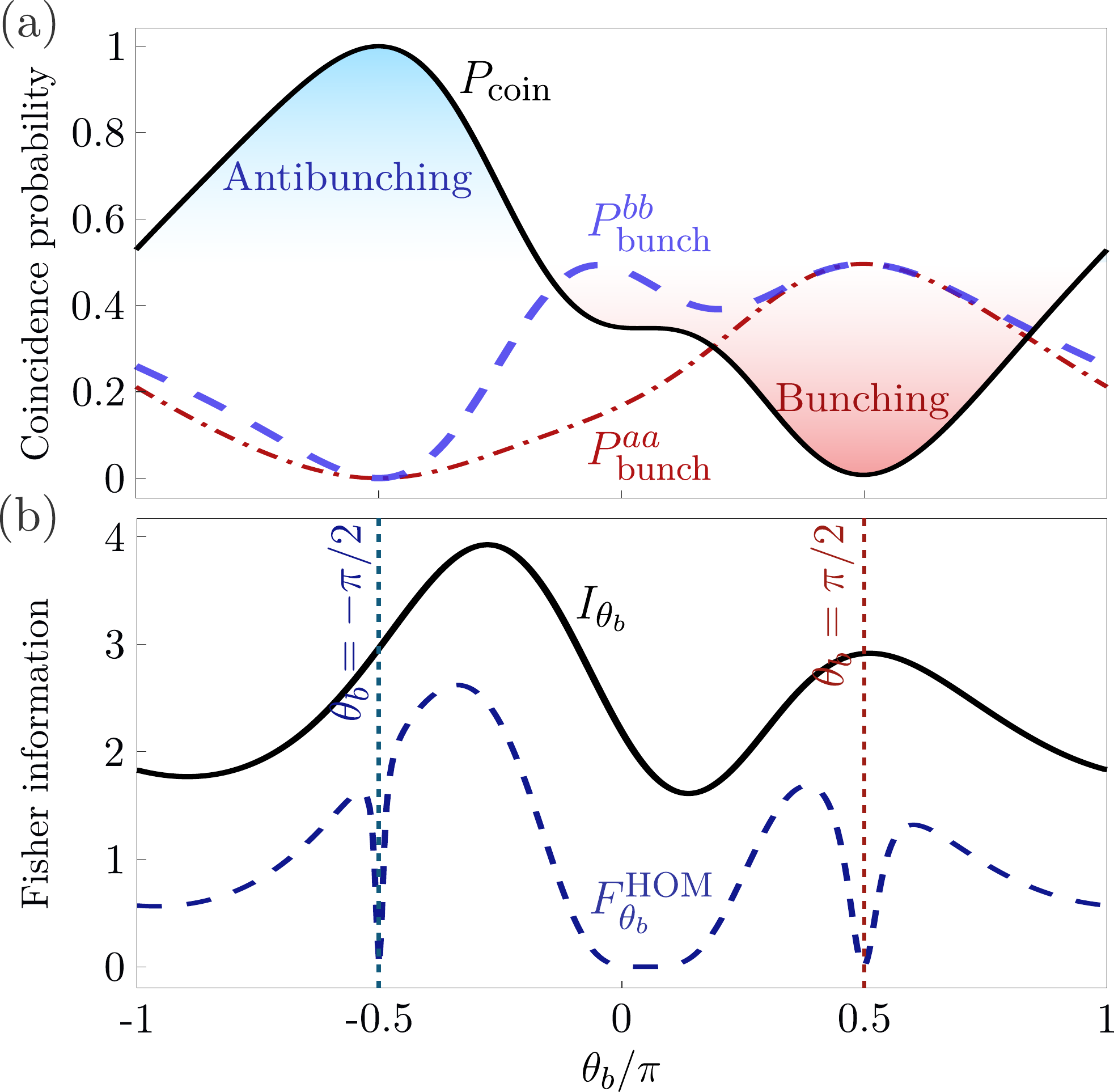}
\caption{Coupling-phase estimation via HOM statistics.
(a) Generalized coincidence probability $P_{\rm coin}$ and the same-port bunching probabilities $P^{aa}_{\rm bunch}$ and $P^{bb}_{\rm bunch}$ as functions of $\theta_b$, with the reference phase fixed at $\theta_a=\pi/2$. The shaded regions indicate bunching- and antibunching-dominated regimes. 
(b) Classical Fisher information $F_{\theta_b}^{\rm HOM}$ (blue dashed) associated with the binary HOM measurement  and quantum Fisher information $I_{\theta_b}$ (black solid) of the asymptotic two-photon output state $\ket{\Phi_{\rm out}(\theta_b)}$. 
Parameters: $N=1$, $w_a=w_b=25$, $x_a=x_b=-50$, $g_a=g_b=J$, $\Delta_e=-2J$, $k_a=k_b=\pi/2$, and $\theta_a=\pi/2$.
\label{fig:thetab-QFI}}
\end{figure}

For an unbiased estimator, $\hat \theta_b$, the fundamental limit to the precision achievable is set by the quantum Cramér--Rao bound ~\cite{PolinoPhotonicQuantum2020, CramerMathematicalMethods1991,RaoInformationAccuracy1992}
\begin{equation}
        \Delta^2 \theta_b \geq \frac{1}{M F^{\text{HOM}}_{\theta_b}} \geq \frac{1}{M I_{\theta_b}},
    \label{eq:QCRB}
\end{equation}
where $M$ is the number of independent repetitions, and $I_{\theta_b}$ is the quantum Fisher information (QFI), defined as the optimization of the CFI over all POVMs. The QFI therefore quantifies the ultimate sensitivity encoded in the output state, independently of the particular measurement employed. For the normalized pure output state considered here, the QFI can be computed via~\cite{ParisQUANTUMESTIMATION2009}
\begin{equation}
I_{\theta_b}=4\mleft[\langle \partial_{\theta_b}\Phi_{\rm out}|\partial_{\theta_b}\Phi_{\rm out}\rangle-\mleft|\langle\Phi_{\rm out}|\partial_{\theta_b}\Phi_{\rm out}\rangle\mright|^2\mright].
\label{eq:QFI}
\end{equation}

Figure~\figpanelNoPrefix{fig:thetab-QFI}{a} shows again how $\theta_b$ is encoded in the two-photon output statistics for the reference phase fixed at $\theta_a=\pi/2$. Varying $\theta_b$ changes both the generalized coincidence probability $P_{\rm coin}$ and the same-port bunching probabilities $P^{aa}_{\mathrm{bunch}}$ and $P^{bb}_{\mathrm{bunch}}$, providing an experimentally accessible, phase-sensitive signal for locally estimating $\theta_b$.

Using the phase-dependent output state $\ket{\Phi_{\rm out}(\theta_b)}$, we evaluate the CFI associated with the binary HOM measurement from \cref{eq:Fi_HOM} and the QFI from \cref{eq:QFI}. The results are shown in \figpanel{fig:thetab-QFI}{b}. We first focus on the region around $\theta_b=\pi/2$, where the coincidence probability is strongly suppressed, as shown in \figpanel{fig:thetab-QFI}{a} and discussed in \secref{sec:HOM}. 

At this point, the CFI (blue dashed) exhibits a pronounced dip and approaches zero, indicating that the HOM minimum is not the optimal operating point for estimating $\theta_b$. Near the minimum, the measured probabilities are only weakly sensitive, to first order, to variations in $\theta_b$. Instead, the most useful metrological information is found in neighboring regions, where small changes in $\theta_b$ produce substantial changes in the output statistics and the CFI reaches values of order unity. Importantly, the QFI (black solid) remains appreciable around $\theta_b=\pi/2$ even when the CFI is substantially suppressed, showing that phase information remains encoded in $\ket{\Phi_{\rm out}(\theta_b)}$ but is not efficiently extracted by simply distinguishing bunching from antibunching. A more general measurement could therefore, in principle, achieve a higher sensitivity. 

Conversely, regions where both the CFI and QFI are suppressed correspond to intrinsically weak phase encoding. Thus, the comparison between the two quantities distinguishes limitations of the specific HOM measurement from the intrinsic phase sensitivity of the scattered two-photon state and highlights the strong dependence of the accessible sensitivity on the operating point.

Implementing the proposed HOM measurement with itinerant microwave photons remains technologically challenging in superconducting circuits~\cite{NarlaRobustConcurrent2016, Gu2017, OpremcakMeasurementSuperconducting2018, KonoQuantumNondemolition2018, BesseSingleShotQuantum2018, AlbertMicrowavePhotonNumber2024}, the primary platform for GA implementations~\cite{GA.Exp, Manenti2017, Andersson2019NP, meandering.Kannan, Vadiraj2021, JoshiResonanceFluorescence2023, Hu2024, Jouanny2025, Xiao2025, AlmanaklyDrivendissipativeEntanglement2026, Yang2026}, although recent experiments have demonstrated substantial progress in microwave photon-counting efficiency~\cite{May2025, Oppliger2026, Kulkarni2026}. The present analysis should therefore be regarded as a theoretical proof of principle for the metrological potential of HOM output statistics in this setting. 

The approach considered here connects with a broader framework of quantum sensing in which information about an unknown parameter is retrieved from the radiation emitted by an open quantum system~\cite{MabuchiInversionQuantum1996, 
GambettaStateDynamical2001,TsangContinuousQuantum2012,GammelmarkBayesianParameter2013,GammelmarkFisherInformation2014,CatanaFisherInformations2015},  with recent applications to parameter estimation and sensing in driven quantum emitters~\cite{AlbarelliUltimateLimits2017,YangEfficientInformation2023,KhanahmadiQubitReadout2023,CabotContinuousSensing2024, GoreckiInterplayTime2025,GoreckiTimeCorrelations2026,CabotParameterEstimation2026,Vivas-VianaTwophotonResonance2021,Alejandro2026}.
%

%
In this context, the gap between the QFI and the information extracted by the binary HOM measurement indicates that measurements resolving additional properties of the scattered field could access phase information discarded by the bunching–antibunching statistics. In superconducting circuits, quadrature measurements provide one experimentally established route for accessing additional information encoded in the output field ~\cite{Gu2017,EichlerCharacterizingQuantum2012,DaSilva2010,Bozyigit2011,Virally2016}.
Complementary metrological strategies within the few-photon scattering setting, based on single-photon scattering and alternative operating points for the two-photon protocol, are discussed in \appref{app:alternative_metrology}.


\section{Conclusion and outlook}
\label{sec:conclusion}

We have shown that a giant atom (GA) with tunable coupling phases can operate as a programmable Hong--Ou--Mandel (HOM) interferometer, with its phase-dependent output statistics additionally providing a resource for coupling-phase estimation. The system we analyzed consists of a single two-level emitter coupled to two waveguides in the form of coupled-cavity arrays (CCAs), each at two separate cavity sites. The coupling-phase differences between spatially separated coupling points continuously reshape the GA self-interference and thereby control its photon-scattering response. This tunability allows the same quantum scatterer to realize balanced beam splitting and programmable two-photon interference.

From the analytical expressions for single-photon scattering, we identified the conditions for reflection-free balanced splitting, a signature of a 50:50 beam splitter, under which a photon incident through either CCA is equally distributed between the two forward output ports. 
Numerical simulations of two incident Gaussian single-photon wave packets then revealed a pronounced HOM dip at this balanced-splitting point. Away from this operating point, tuning the coupling phases continuously changes the correlated two-photon output from bunching into the same port to antibunching across distinct output ports, demonstrating the programmable character of the GA HOM interferometer.
Finally, by analyzing the classical Fisher information associated with a binary HOM measurement and the quantum Fisher information of the scattered state, we showed that the phase-dependent output statistics can be exploited for coupling-phase estimation, revealing the role of both the operating point and the measurement choice in determining the achievable sensitivity.

These results demonstrate the potential of GAs as programmable few-photon scatterers and suggest a route toward reconfigurable two-photon interferometry, correlation engineering, and phase-sensitive measurements in waveguide-QED platforms. The ability to control beam splitting and two-photon correlations within the same quantum scatterer may provide useful building blocks for integrated quantum interconnects and dynamically configurable photonic quantum devices, with applications in quantum information processing and quantum communication. Extending this approach to multiple GAs, more complex waveguide geometries, or higher-photon number inputs could enable richer forms of programmable quantum transport and multi-photon interference.


\acknowledgments
YL acknowledges support from the Quantum Science and Technology-National Science and Technology Major Project (Grant No.~2023ZD0300704) and the National Natural Science Foundation of China (Grant Nos.~12274107, 12574387, and 12547103). LD acknowledges financial support from the Knut och Alice Wallenberg stiftelse through project grant No.~2022.0090.
AVV and AFK acknowledge support from the Swedish Foundation for Strategic Research (Grant No.~FFL21-0279). AFK is also supported by the Swedish Foundation for Strategic Research (Grant No.~FUS21-0063), the Horizon Europe programme HORIZON-CL4-2022-QUANTUM-01-SGA via the project 101113946 OpenSuperQPlus100, the Norwegian Research Council through the Norwegian Quantum Software Center (NorQSoft, project number 361350), and the Knut and Alice Wallenberg Foundation through the Wallenberg Centre for Quantum Technology (WACQT).


\appendix


\section{Self-energy for a giant atom coupled to two independent coupled-cavity arrays}
\label{app:self_energy_two_lattices}

Here, we compute the self-energy of a giant atom coupled to two CCAs using the resolvent formalism~\cite{Cohen-TannoudjiAtomPhotonInteractions1998}, adapting the method used in Ref.~\cite{Soro_2023} to the present setup.

Let $\hat Q$ denote the projector onto the single-photon bath subspace of the two CCAs, excluding the atomic excited state. Since the one-photon states in the $a$ and $b$ CCAs belong to orthogonal sectors, the bath projector can be decomposed as
\begin{equation}
\hat Q = \hat Q_a + \hat Q_b,
\end{equation}
where each subprojector is defined as
\begin{equation}
\hat Q_\alpha\equiv
\sum_m\hat{\alpha}_m^\dagger\ket{0,g}\bra{0,g}\hat{\alpha}_m,
\qquad \alpha=a,b,
\end{equation}
and satisfies
\begin{equation}
\hat Q_\alpha \hat Q_\beta=\delta_{\alpha\beta}\hat Q_\alpha,
\qquad \alpha,\beta=a,b.
\end{equation}
For the momentum-space formulation, we temporarily regularize each infinite CCA by considering a periodic lattice containing $L$ sites. The normalized momentum-space single-photon states are
\begin{equation}
\ket{k,\alpha}
=\frac{1}{\sqrt{L}}\sum_m e^{ikm}
\hat\alpha_m^\dagger\ket{0,g}.
\end{equation}

Within the subspace onto which $\hat Q$ projects, the atom--photon coupling terms do not contribute to $\hat Q\hat H\hat Q$, because they only connect the atomic excited-state subspace to the single-photon bath subspace. 
Furthermore, the system Hamiltonian in~\cref{eq:H_int} does not contain any direct hopping or conversion term between the two CCAs. 
Thus,
\begin{equation}
\hat Q_a  \hat H\hat Q_b = \hat Q_b \hat H \hat Q_a =0,
\end{equation}
and hence
\begin{equation}
 \hat Q\hat H\hat Q=\hat Q_a\hat  H \hat Q_a+\hat Q_b  \hat H \hat Q_b .
\end{equation}
The restricted resolvent is therefore block-diagonal,
\begin{equation}
    \frac{1}{z-\hat  Q\hat H\hat Q}=
\frac{1}{z-\hat  Q_a\hat H\hat Q_a-\hat Q_b\hat H\hat Q_b}.
\end{equation}

An eigenstate in the $a$ CCA satisfies $\hat Q_a\ket{k,a}=\ket{k,a}$ and $\hat Q_b\ket{k,a}=0$. Therefore,
\begin{equation}
    \frac{1}{z-\hat Q\hat H\hat Q}\ket{k,a}
= \frac{1}{z-\hat Q_a\hat H\hat Q_a}\ket{k,a} = \frac{1}{z-\omega(k)}\ket{k,a}.
\end{equation}
Similarly, an eigenstate in the $b$ CCA satisfies $\hat Q_b\ket{k,b}=\ket{k,b}$ and $\hat Q_a\ket{k,b}=0$, then
\begin{equation}
    \frac{1}{z-\hat Q\hat H\hat Q}\ket{k,b}
= \frac{1}{z-\hat Q_b\hat H\hat Q_b}\ket{k,b} = \frac{1}{z-\omega(k)}\ket{k,b}.
\end{equation}

Consequently, the excited-state self-energy in the two-CCA model can be decomposed into independent contributions from the two CCAs. Denoting the atom–CCA interaction part of \cref{eq:H_int} by $\hat V$, we define
\begin{equation}
\Sigma_e(z)=\bra{0,e}\hat V\frac{1}{z-\hat  Q\hat H\hat Q}\hat V\ket{0,e},
\end{equation}
which can be decomposed into two independent channel contributions,
\begin{equation}
\Sigma_e(z)=\Sigma_a(z)+\Sigma_b(z),
\end{equation}
with
\begin{equation}
\Sigma_\alpha(z)\equiv\sum_k \frac{\mleft|\bra{k,\alpha}\hat V\ket{0,e}\mright|^2}{z-\omega(k)},\quad \alpha=a,b.
\end{equation}
In the present model, the tight-binding dispersion relation is 
\begin{equation}
\omega(k)=-2J\cos k,
\end{equation}
and the giant-atom coupling matrix element to $\alpha$ CCA in momentum space is
\begin{equation}
V_{k,\alpha}=\bra{k,\alpha}\hat V\ket{0,e}=\frac{g_\alpha}{\sqrt{L}}\mleft[1+e^{i(\theta_\alpha-kN)}\mright].
\end{equation}

The single-CCA self-energy contribution thus becomes
\begin{equation}
\Sigma_\alpha(z)=\frac{2g_\alpha^2}{L}\sum_k\frac{1+\cos(\theta_\alpha-kN)}{z+2J\cos k}.
\end{equation}
In the continuum limit $L\to\infty$, the momentum sum is replaced according to $\sum_k\to \frac{L}{2\pi}\int_{-\pi}^{\pi}dk$, and the self-energy becomes
\begin{equation}
\Sigma_\alpha(z)=\frac{g_\alpha^2}{\pi}\int_{-\pi}^{\pi}dk\frac{1+\cos(\theta_\alpha-kN)}{z+2J\cos k}.
\end{equation}
Using $\cos(\theta_\alpha-kN)=\cos\theta_\alpha\cos(kN)+\sin\theta_\alpha\sin(kN)$ and noting that the denominator is even in $k$, whereas $\sin(kN)$ is odd, we have
\begin{equation}
\int_{-\pi}^{\pi}\frac{\sin(kN)}{z+2J\cos k}\,dk=0,
\end{equation}
and therefore
\begin{multline}
    \Sigma_\alpha(z)=\frac{g_\alpha^2}{\pi}  \mleft[\int_{-\pi}^{\pi}\frac{dk}{z+2J\cos k} \mright.\\
+ \mleft. \cos\theta_\alpha\int_{-\pi}^{\pi}\frac{\cos(kN)\,dk}{z+2J\cos k}\mright].
\label{eq:self-energy-formal}
\end{multline}
Moreover, since $e^{ikN}=\cos(kN)+i\sin(kN)$ and the $\sin(kN)$ term integrates to zero, the second integral can equivalently be written as
\begin{equation}
\int_{-\pi}^{\pi}\frac{\cos(kN)\,dk}{z+2J\cos k}=\int_{-\pi}^{\pi}\frac{e^{ikN}\,dk}{z+2J\cos k}.
\end{equation}

The two integrals in \cref{eq:self-energy-formal} can be evaluated by contour integration using the substitution $\zeta=e^{ik}$, for which $dk=d\zeta/(i\zeta)$ and $\cos k=(\zeta+\zeta^{-1})/2$. Then
\begin{equation}
z+2J\cos k=\frac{J\zeta^2+z\zeta+J}{\zeta}.
\end{equation}
The first integral therefore becomes
\begin{equation}
\int_{-\pi}^{\pi}\frac{dk}{z+2J\cos k}=\frac{1}{i}\oint_{|\zeta|=1}\frac{d\zeta}{J\zeta^2+z\zeta+J}.
\end{equation}
The poles are the roots of $J\zeta^2+z\zeta+J=0$,
\begin{equation}
f_\pm(z)=\frac{-z\pm\sqrt{z^2-4J^2}}{2J},\qquad f_+(z)f_-(z)=1.
\end{equation}
For the retarded self-energy, we take $\operatorname{Im}z>0$ and choose the branch of $\sqrt{z^2-4J^2}$ that is analytic in the upper half-plane and satisfies $\sqrt{z^2-4J^2}\sim z$ as $|z|\to\infty$. With this branch choice, the root $f_+(z)$ lies inside the unit circle, whereas $f_-(z)=1/f_+(z)$ lies outside. The residue theorem then yields
\begin{equation}
\int_{-\pi}^{\pi}\frac{dk}{z+2J\cos k}=\frac{2\pi}{\sqrt{z^2-4J^2}}.
\end{equation}
Similarly, the second integral becomes%
\begin{equation}
\int_{-\pi}^{\pi}\frac{e^{ikN}\,dk}{z+2J\cos k}=\frac{1}{iJ}\oint_{|\zeta|=1}\frac{\zeta^N\,d\zeta}{(\zeta-f_+)(\zeta-f_-)}.
\end{equation}
Evaluating the residue at $f_+(z)$, the pole inside $|\zeta|<1$, gives
\begin{equation}
\int_{-\pi}^{\pi}\frac{e^{ikN}\,dk}{z+2J\cos k}=\frac{2\pi}{\sqrt{z^2-4J^2}}
\mleft(\frac{-z+\sqrt{z^2-4J^2}}{2J}\mright)^N.
\end{equation}
Substituting these results into \cref{eq:self-energy-formal} yields
\begin{equation}
\Sigma_\alpha(z)=\frac{2g_\alpha^2}{\sqrt{z^2-4J^2}}\mleft[1+\cos\theta_\alpha\mleft(\frac{-z+\sqrt{z^2-4J^2}}{2J}\mright)^N\mright].
\end{equation}
For the on-shell retarded self-energy, we set $z=E_k+i0^+$, where
$E_k=-2J\cos k$ and $0<k<\pi$. We then obtain
\begin{equation}
\Sigma_\alpha(E_k+i0^+)=\frac{g_\alpha^2}{iJ\sin k}\mleft(1+e^{ikN}\cos\theta_\alpha\mright).
\label{eq:onshellenergy}
\end{equation}
The quantity $D_\alpha$ introduced in \cref{eq:Dalpha} can therefore be identified with the on-shell retarded self-energy, $D_\alpha=\Sigma_\alpha(E_k+i0^+)$. The total self-energy is therefore $\Sigma_e=\Sigma_a+\Sigma_b$. We define the total radiative decay rate $\Gamma(E_k)$ through
\begin{equation}
\Sigma_e(E_k+i0^+)=\mathrm{Re}\,\Sigma_e(E_k+i0^+)-\frac{i}{2}\Gamma(E_k).
\end{equation}
The decay rate into the $\alpha$ CCA is
\begin{equation}
\Gamma_\alpha(E_k)=-2\,\mathrm{Im}\,\Sigma_\alpha(E_k+i0^+).
\end{equation}
At the working point considered in the main text, namely $k=\pi/2$, $N=1$, $\theta_a=\theta_b=\pi/2$, and $g_a=g_b=g$, \cref{eq:onshellenergy} reduces to
\begin{equation}
\Sigma_\alpha(E_k+i0^+) = -\frac{ig^2}{J},
\quad \alpha=a,b.
\end{equation}
Consequently, the radiative decay rate into each CCA is
\begin{equation}
\Gamma_\alpha = \frac{2g^2}{J},
\quad \alpha=a,b.
\end{equation}
Since the two CCAs constitute independent decay channels, the total radiative decay rate is
\begin{equation}
\Gamma=\Gamma_a+\Gamma_b
=
\frac{4g^2}{J}.
\end{equation}
This total decay rate determines the full linewidth of the emitter. At the working point considered here, the reflection-free balanced-splitting condition requires the detuning magnitude to equal this half-width, yielding
\begin{equation}
\Delta_e
=\pm \frac{\Gamma}{2}=
\pm\frac{2g^2}{J}.
\end{equation}
%

\section{Two-excitation equations of motion and initial-state construction}
\label{app:eom}

In this appendix, we present the details of the two-excitation calculations used in \secref{sec:TwoPhotonScattering}. We first show how the incident two-photon wave packet is constructed from two single-photon wavepacket operators and then derive the coupled equations of motion for the amplitudes in \cref{eq:TwoPhotonState}.

To construct the initial state in the two-excitation subspace, we introduce two single-excitation creation operators,
\begin{equation}
\hat \phi^{(q)\dagger}(0)=\sum_m\mleft[A_m^{(q)}(0)\hat a_m^\dagger+B_m^{(q)}(0)\hat b_m^\dagger\mright]+C_e^{(q)}(0)\sigma^\dagger, 
\end{equation}
with $q=1,2$. Here, $A_m^{(q)}(0)$ and $B_m^{(q)}(0)$ are the single-photon amplitudes of the $q$th wave packet at site $m$ of the $a$ and $b$ CCAs, respectively, while $C_e^{(q)}(0)$ is the corresponding atomic excitation amplitude. The initial two-excitation state is then constructed as
\begin{equation}
\ket{\Phi_2(0)}=\mathcal{N}_{12}\hat\phi^{(1)\dagger}(0)\hat\phi^{(2)\dagger}(0)\ket{0,g},
\end{equation}
where $\mathcal{N}_{12}$ is a normalization constant. Expanding this product in the basis introduced in \cref{eq:TwoPhotonState} and collecting the contributions according to bosonic symmetry yields the following initial amplitudes, up to the common normalization factor
\begin{subequations}
    \begin{align}
    C_{m,n}^{aa}(0) &\propto \mleft[A_{m}^{(1)}(0)A_{n}^{(2)}(0) + A_{n}^{(1)}(0)A_{m}^{(2)}(0)\mright]/\sqrt{1+\delta_{m,n}}, \\
    C_{m,n}^{bb}(0) &\propto \mleft[B_{m}^{(1)}(0)B_{n}^{(2)}(0) + B_{n}^{(1)}(0)B_{m}^{(2)}(0)\mright]/\sqrt{1+\delta_{m,n}}, \\
    C_{m,n}^{ab}(0) &\propto A_{m}^{(1)}(0)B_{n}^{(2)}(0) + A_{m}^{(2)}(0)B_{n}^{(1)}(0), \\
    C_{m}^{ea}(0) &\propto A_{m}^{(1)}(0)C_{e}^{(2)}(0) + A_{m}^{(2)}(0)C_{e}^{(1)}(0), \\
    C_{m}^{eb}(0) &\propto B_{m}^{(1)}(0)C_{e}^{(2)}(0) + B_{m}^{(2)}(0)C_{e}^{(1)}(0).
    \end{align}
\end{subequations}
No doubly excited atomic component appears because the GA is a two-level system, for which  $(\hat\sigma^\dagger)^2=0$.

Substituting the state expansion in \cref{eq:TwoPhotonState} into the time-dependent Schr\"odinger equation yields the following coupled equations of motion for the probability amplitudes:
\begin{widetext}
\begin{subequations}
    \begin{align}
    i\frac{dC_{m,n}^{\alpha\alpha}}{dt} =& -J\sqrt{1+\delta_{m,n}}\mleft(C_{m+1,n}^{\alpha\alpha}+C_{m,n-1}^{\alpha\alpha}\mright)-J\mleft(1-\delta_{m,n}\mright)\sqrt{1+\delta_{m,n+1}}\mleft(C_{m-1,n}^{\alpha\alpha}    +C_{m,n+1}^{\alpha\alpha}\mright)\notag\\
    & +\frac{g_\alpha}{\sqrt{1+\delta_{m,n}}}\mleft[\mleft(\delta_{m,0}+e^{i\theta_\alpha}\delta_{m,N}\mright)C_n^{e\alpha}+\mleft(\delta_{n,0}+e^{i\theta_\alpha}\delta_{n,N}\mright)C_m^{e\alpha}\mright],\qquad m\geq n,\quad \alpha=a,b.\label{eq:B3a}\\
    i\frac{dC_{m,n}^{ab}}{dt} =& -J\mleft(C_{m+1,n}^{ab}+C_{m-1,n}^{ab}+C_{m,n+1}^{ab}+C_{m,n-1}^{ab}\mright)+g_{b}\mleft(\delta_{n,0}C_{m}^{ea}+e^{i\theta_b}\delta_{n,N}C_{m}^{ea}\mright)+g_{a}\mleft(\delta_{m,0}C_{n}^{eb}+e^{i\theta_a}\delta_{m,N}C_{n}^{eb}\mright),\label{eq:B3b}\\
    i\frac{dC_{m}^{ea}}{dt} =& \Delta_{e}C_{m}^{ea}-J\mleft(C_{m+1}^{ea}+C_{m-1}^{ea}\mright)+g_a\left[\sqrt{1+\delta_{m,0}}\,C_{m,0}^{aa}+e^{-i\theta_a}\sqrt{1+\delta_{m,N}}\,C_{m,N}^{aa}\right]+g_{b}\mleft(C_{m,0}^{ab}+e^{-i\theta_b}C_{m,N}^{ab}\mright),\label{eq:B3c}\\
    i\frac{dC_{m}^{eb}}{dt} =& \Delta_{e}C_{m}^{eb}-J\mleft(C_{m+1}^{eb}+C_{m-1}^{eb}\mright)+g_b\left[\sqrt{1+\delta_{m,0}}\,C_{m,0}^{bb}+e^{-i\theta_b}\sqrt{1+\delta_{m,N}}\,C_{m,N}^{bb}\right]+g_{a}\mleft(C_{0,m}^{ab}+e^{-i\theta_a}C_{N,m}^{ab}\mright).\label{eq:B3d}
    \end{align}
\end{subequations}
\end{widetext}
In the $C_{m,n}^{aa}$ and $C_{m,n}^{bb}$ sectors, only the independent amplitudes with ordered indices $m\geq n$ are retained. \Cref{eq:B3a} applies throughout this ordered domain, including both off-diagonal and diagonal cases. For $m>n+1$, it reduces to the usual off-diagonal equation of motion. For $m=n+1$, the factor $\sqrt{1+\delta_{m,n+1}}$ accounts for the bosonic enhancement of the coupling to the diagonal amplitudes. For $m=n$, the factor $1-\delta_{m,n}$ removes the duplicated terms, while the remaining hopping and atom--photon coupling terms acquire the required factor $\sqrt{2}$. Any same-CCA amplitude appearing with unordered indices in ~\cref{eq:B3c,eq:B3d} are rewritten in the ordered form using bosonic exchange symmetry. Therefore,~\crefrange{eq:B3a}{eq:B3d} form a closed set of equations of motion, so that no separate equations for the diagonal amplitudes are required.


\section{Dependence on the coupling-point separation}
\label{app:alternative_separation}

In the main text, we focus on the case $N=1$ (i.e., the GA couples to two adjacent cavities in each CCA) to illustrate the basic mechanism of coupling-phase-controlled single-photon scattering and two-photon interference.  Here, we show that the observed behavior is not specific to this choice of coupling-point separation, but instead originates from the general self-interference mechanism of a giant atom. For a photon with wave vector $k$, the relative phase between the coupling points combines the propagation phase $kN$ with the externally controlled coupling phase. Therefore, changing $N$ modifies the effective interference condition without changing the underlying mechanism.

\begin{figure}[b!]
\includegraphics[width=\linewidth]{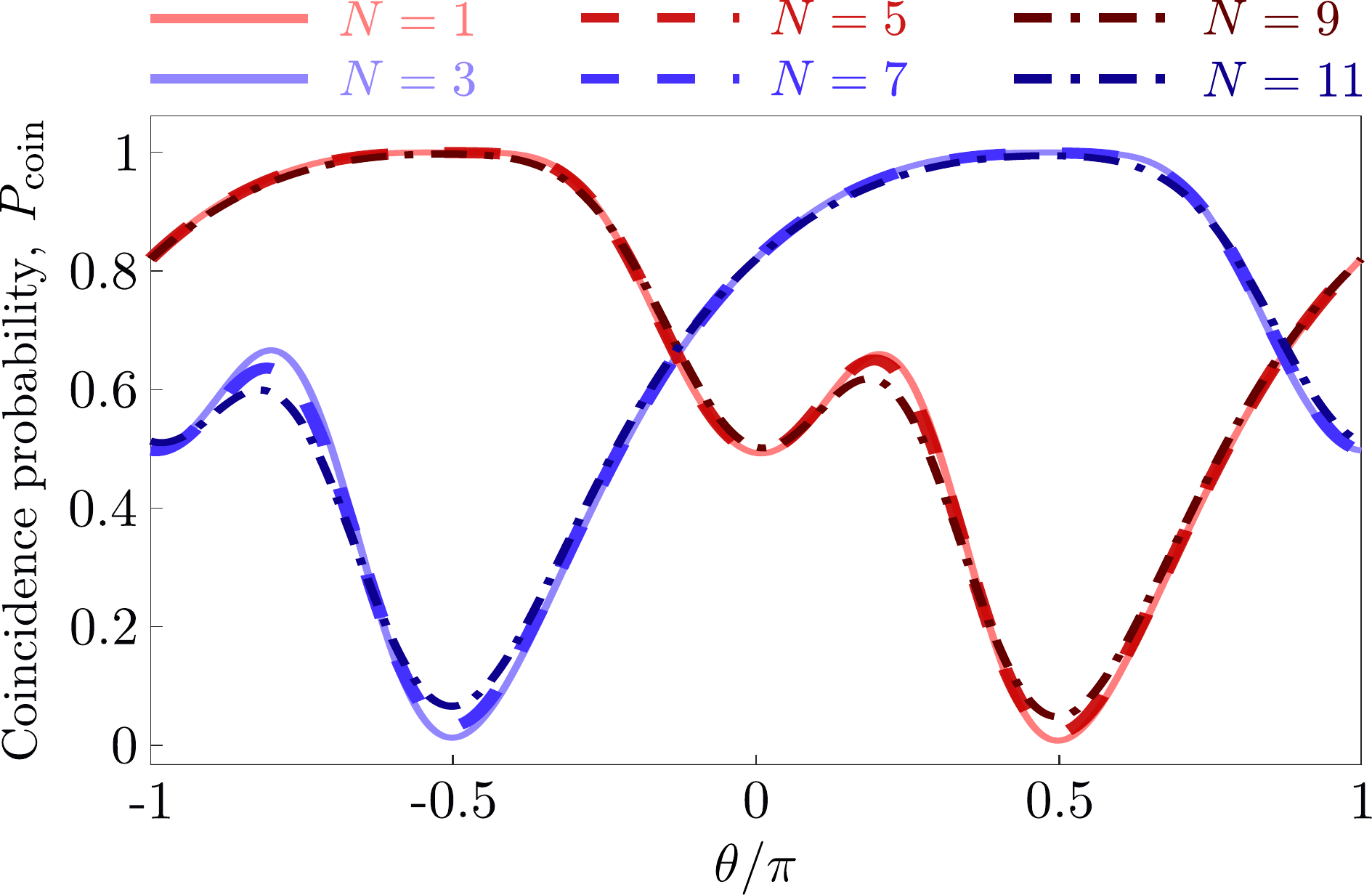}
\caption{Phase dependence of the two-photon coincidence probability for different odd coupling-point separations $N$. The synchronized coupling phases are varied as $\theta_a=\theta_b\equiv\theta$. Red and blue curves correspond to the $N=4\ell+1$ and $N=4\ell+3$ families, respectively. 
Parameters: $w_a=w_b=25$, $x_a=x_b=-50$, $g_a=g_b=J$, $k_a=k_b=\pi/2$, $\Delta_e=-2J$.
\label{fig:DiffN}}
\end{figure}

Figure~\ref{fig:DiffN} shows that the phase-dependent suppression and enhancement of the coincidence probability persist as the coupling-point separation is increased. Compared with \figpanel{fig:HOM-dip-theta}{c} in the main text for $N = 1$,  the same qualitative phase-controlled two-photon interference is observed for all the odd values of $N$ considered here, confirming that this behavior is not specific to adjacent coupling points. 

The change of the interference pattern with $N$ can be understood from the effective accumulated phase between the two coupling points, which contains both the propagation phase $kN$ and the externally controlled phase $\theta$. At the resonant wave vector $k=\pi/2$, the odd coupling-point separations naturally separate into two families, $N=4\ell+1$ and $N=4\ell+3$, with $\ell\in\mathbb{N}_0$. Separations within each family accumulate the same propagation phase modulo $2\pi$, whereas the two families differ by a phase shift of $\pi$. Hence, the curves for $N = 1, 5, 9$ in \figref{fig:DiffN} are very similar, while their phase dependence is shifted by approximately $\pi$ relative to that of the curves for $N = 3, 7, 11$. For even $N$, the propagation phase becomes an integer multiple of $\pi$, so the directional interference required for the reflection-free balanced-splitting condition cannot be realized at $k=\pi/2$. We therefore restrict to odd $N$.

In \figref{fig:DiffN}, we also observe a small reduction in the depth of the phase-controlled HOM dip as the coupling-point separation increases within each family. This behavior originates from the finite spectral width of the incident wave packets. 
In fact, the propagation contribution $kN$ to the accumulated phase varies across the finite momentum bandwidth of the incident photons. As $N$ increases, this phase variation becomes larger, so that different spectral components experience slightly different interference conditions. The resulting spectral averaging weakens the HOM interference contrast. For this reason, we choose the minimal separation $N=1$ in the main text, for which these finite-bandwidth effects are minimized and the phase-controlled HOM interference is most clearly displayed.


\section{Additional parameter estimation strategies}
\label{app:alternative_metrology}

In this appendix, we expand on the quantum-metrology applications of our GA beam splitter discussed in \secref{sec:metrology}. We first consider phase estimation from single-photon scattering and then extend the two-photon analysis by relaxing the fixed-$\theta_a$ condition imposed in the main text.


\begin{figure}[t!]
\includegraphics[width=\linewidth]{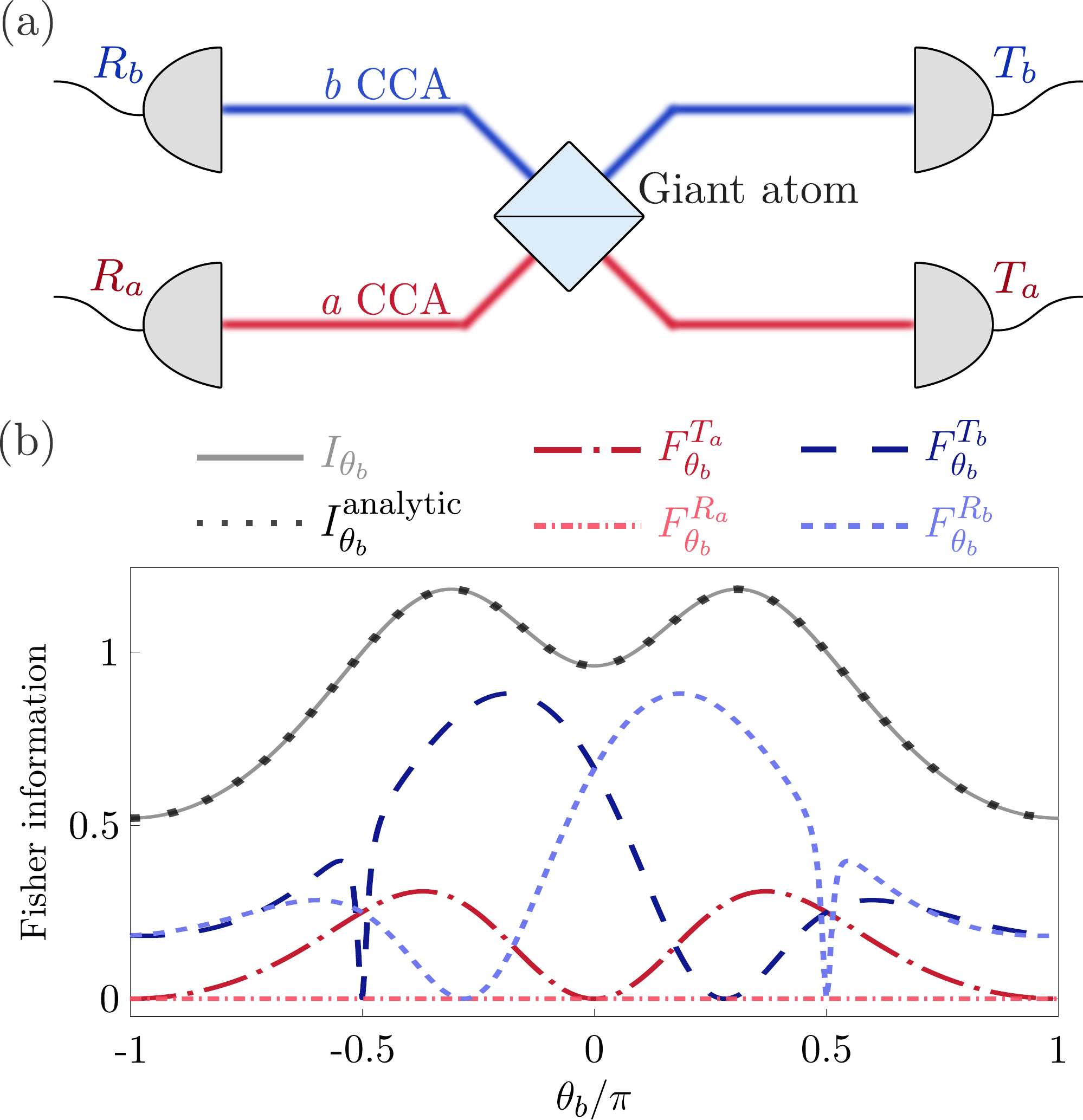}
\caption{Coupling-phase estimation via single-photon scattering.
(a) Schematic of the measurement setup. 
(b) Classical Fisher information $F_{\theta_b}^{\nu}$ associated with binary detection at a selected output port $\nu$, with the other three ports grouped into the complementary outcome, and quantum Fisher information $I_{\theta_b}$ of the asymptotic single-photon output state. The gray solid curve shows the numerical QFI $I_{\theta_b}$, while the gray dotted curve shows the analytical result $I_{\theta_b}^{\mathrm{analytic}}$.
The red dash-dotted and light-red dash-dotted curves show $F_{\theta_b}^{T_a}$ and $F_{\theta_b}^{R_a}$, respectively, while the dark- and light-blue dashed curves show $F_{\theta_b}^{T_b}$ and $F_{\theta_b}^{R_b}$, respectively.
Parameters: $N=1$, $k=\pi/2$, $g_a=g_b=J$, $\Delta_e=-2J$, $\theta_a=\pi/2$, $x=-50$, and $w=25$. 
\label{fig:A1}}
\end{figure}

\subsection{Single-photon scattering}

We now analyze the phase estimation problem using the single-photon scattering process derived in \secref{sec:SinglePhotonScattering}.
For a photon incident from the left of the $a$ CCA, the scattering process is characterized by the transmission amplitudes $t_a$ and $t_b$ and the reflection amplitudes $r_a$ and $r_b$.
The expressions for these amplitudes are given in~\crefrange{eq:ta}{eq:rb}.

Just like for two-photon scattering in \secref{sec:metrology}, we consider the case where $\theta_a$ is fixed while $\theta_b$ varies and is treated as an unknown parameter. For the metrological analysis, we focus on the asymptotic output after the scattering process. 
The amplitudes inside the scattering region and the atomic excitation amplitude enter the scattering dynamics but do not constitute independent asymptotic output channels. In the absence of intrinsic loss, the single-photon scattering amplitudes satisfy the flux conservation relation 
\begin{equation}
|t_a|^2+|t_b|^2+|r_a|^2+|r_b|^2=1 .
\end{equation}
Therefore, for a fixed incident wave vector, the normalized asymptotic output state can be written in terms of the four output ports as
\begin{multline}
    \ket{\psi_{\rm out}(\theta_b)} \equiv t_a(\theta_b)\ket{T_a}+t_b(\theta_b)\ket{T_b}  \\ +r_a(\theta_b)\ket{R_a}+r_b(\theta_b)\ket{R_b}.
\label{asymptotic}
\end{multline}
Here, $\ket{T_a}$ and $\ket{T_b}$ denote the transmitted single-photon in the two CCAs, while $\ket{R_a}$ and $\ket{R_b}$ denote the corresponding reflected states.
These channel states are mutually orthogonal and independent of the coupling phase $\theta_b$.

Using the expression for the quantum Fisher information already presented in \cref{eq:QFI}, the single-photon QFI can be written directly in terms of the complex scattering amplitudes as
\begin{equation}
I^{\rm analytic}_{\theta_b}=4\mleft[\sum_{\mu}\mleft|\partial_{\theta_b} \mu\mright|^2-\left|\sum_{\mu}\mu^*\partial_{\theta_b} \mu\right|^2\mright],
\label{eq:QFIana}
\end{equation}
where $\mu=\{t_a,t_b,r_a,r_b\}$.
This expression shows that the single-photon QFI is determined by the $\theta_b$-dependence of the full complex scattering amplitudes, including both their magnitudes and phases. Since analytical expressions for the scattering amplitudes are available, the same expression can be readily generalized to other system parameters by taking the corresponding parameter derivatives.

Using the analytical scattering coefficients obtained in~\crefrange{eq:ta}{eq:rb} for $N=1$, $k=\pi/2$, and $g_a=g_b\equiv g$, we obtain a closed-form expression
\begin{multline}
    I^{\rm analytic}_{\theta_b}=\frac{4\left(1+\sin \theta_a\right)} {\left(R^2+4\right)^2} \left[R^2+2\left(1-\sin \theta_a\right) \sin \theta_b \right. \\
    \left. +\left(3-\sin \theta_a\right)\left(1+\sin ^2 \theta_b\right)\right],
    \label{eq:QFI-analytic}
\end{multline}
with $R=J \Delta_e/g^2+\cos \theta_a+\cos \theta_b$. This analytical result provides a benchmark for comparison with the classical Fisher information obtained from output-port detection in the numerical wave-packet simulations below.

\begin{figure}[t!]
\includegraphics[width=1\linewidth]{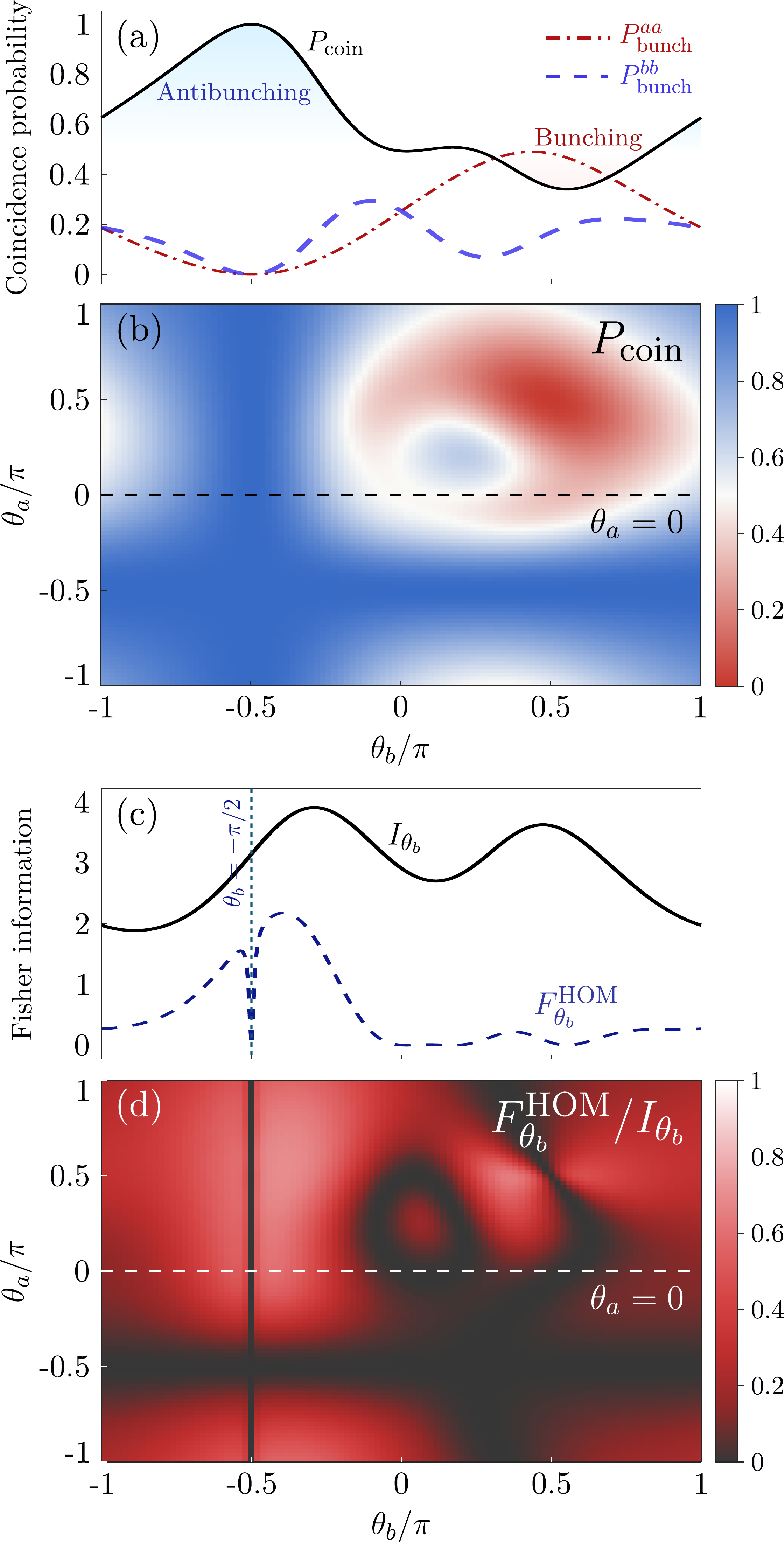}
\caption{Coupling-phase estimation via HOM statistics in the ($\theta_a,\theta_b$) space.
(a) Generalized coincidence probability $P_{\rm coin}$ and same-port bunching probabilities $P_{\rm bunch}^{aa}$ and $P_{\rm bunch}^{bb}$ as functions of $\theta_b$ for $\theta_a=0$.
(b) Generalized coincidence probability $P_{\rm coin}$ as a function of both coupling phases $\theta_a$ and $\theta_b$. The black dashed line at $\theta_a=0$ indicates the one-dimensional cut shown in (a).
(c) Classical Fisher information $F_{\theta_b}^{\rm HOM}$ (blue dashed) associated with the binary HOM measurement and quantum Fisher information $I_{\theta_b}$ (black solid)  as functions of $\theta_b$ for $\theta_a=0$.
(d) Ratio $F_{\theta_b}^{\rm HOM}/I_{\theta_b}$ (CFI/QFI) as a function of $\theta_a$ and $\theta_b$. The white dashed line at $\theta_a=0$ indicates the one-dimensional cut shown in (c).
Parameters: $N=1$, $w_a=w_b=25$, $g_a=g_b=J$, $x_a=x_b=-50$, $\Delta_e=-2J$, and $k_a=k_b=\pi/2$.
\label{fig:HOM-metrology-thetaab}
}
\end{figure}

In the numerical simulations, a single photon is initialized in the $a$ CCA as a Gaussian wave packet, with initial center $x$, carrier wave vector $k$, and spatial width $w$. We then evolve the state using the wave-packet time-evolution method described for two-photon scattering in \secref{sec:TwoPhotonScattering}. At a sufficiently long final time $t_f$, the scattered wave packet has left the coupling region and the atomic excitation is negligible. We therefore approximate the asymptotic numerical output state by the time-evolved state at $t_f$:
\begin{equation}
\ket{\psi_{\rm out}(\theta_b)}\approx \ket{\Phi_1(t_f;\theta_b)}.
\end{equation}
Accordingly,  the output state is decomposed into the four spatially separated outgoing components
\begin{align}
\ket{\psi_{\rm out}(\theta_b)} &= \sum_{m=N}^{\infty} A_m(t_f;\theta_b)\hat a_m^\dagger\ket{0,g} \nonumber\\
&+\sum_{m=N}^{\infty} B_m(t_f;\theta_b)\hat b_m^\dagger\ket{0,g} \nonumber \\
&+\sum_{m=-\infty}^{0} A_m(t_f;\theta_b)\hat a_m^\dagger\ket{0,g} \nonumber \\
&+\sum_{m=-\infty}^{0} B_m(t_f;\theta_b)\hat b_m^\dagger\ket{0,g} .
\label{Eq:numsingleoutput}
\end{align}
The four terms correspond the transmitted and reflected wave-packet components in the two CCAs. For the detection scheme shown in \figpanel{fig:A1}{a}, the corresponding output-port probabilities are obtained from the norms of these components as
\begin{subequations}
\begin{align}
    P_{T_a}(\theta_b)&=\sum_{m=N}^{\infty} \mleft|A_m(t_f;\theta_b) \mright|^2, \\
    P_{T_b}(\theta_b)&=\sum_{m=N}^{\infty} \mleft|B_m(t_f;\theta_b) \mright|^2 , \\
    P_{R_a}(\theta_b)&=\sum_{m=-\infty}^{0} \mleft|A_m(t_f;\theta_b) \mright|^2 , \\
    P_{R_b}(\theta_b)&=\sum_{m=-\infty}^{0} \mleft|B_m(t_f;\theta_b) \mright|^2 .
\end{align}
\end{subequations}

For each output port, single-photon detection defines a binary measurement: the photon is either detected in the chosen port or not. Therefore, for a given port $\nu\in\{T_a,T_b,R_a,R_b\}$, the corresponding classical Fisher information is defined as
\begin{equation}
    F^{\nu}_{\theta_b}\equiv\frac{\mleft[ \partial_{\theta_b}P_{\nu}(\theta_b)\mright]^2}{P_{\nu}(\theta_b)\mleft[1-P_{\nu}(\theta_b)\mright]}.
    \label{Eq.single.num.CFI}
\end{equation}
In the numerical calculation, $\partial_{\theta_b}P_{\nu}$ is evaluated using finite-difference method from wave-packet evolutions at nearby values of $\theta_b$.


In Fig.~\figpanelNoPrefix{fig:A1}{b}, we compare the numerical binary CFI for each output port, $F_{\theta_b}^\nu$, with the single-photon QFI, $I_{\theta_b}$, noting that the analytical expression presented in Eq.~\eqref{eq:QFI-analytic} perfectly matches with numerical simulations. In contrast, each independent $F^{\nu}_{\theta_b}$ quantifies the sensitivity accessible by distinguishing detection at the selected port $\nu$ from detection at the other three ports. These binary CFIs are strongly port dependent, reflecting how the phase information is redistributed among the four scattering channels as $\theta_b$ varies. For each value of $\theta_b$, the largest $F_{\theta_b}^{\nu}$  therefore identifies the most informative port for single-port binary phase estimation within this detection scheme.

\subsection{Tunable working points for coupling-phase estimation via HOM statistics}

In \secref{sec:metrology}, the phase-estimation protocol using the binary HOM measurement was illustrated by fixing the reference phase of the $a$ CCA to $\theta_a=\pi/2$ and treating $\theta_b$ as the unknown parameter. To examine the dependence of the metrological response on $\theta_a$, we repeat the same two-photon scattering calculation while varying $\theta_a$ from $-\pi$ to $\pi$. For each value of $\theta_a$, the asymptotic two-photon output state $|\Phi_{\rm out}(\theta_b;\theta_a)\rangle$ is obtained from the wave-packet evolution. The generalized coincidence probability $P_{\rm coin}$ is evaluated using the projector defined in \cref{eq.Pcoin}, while the quantum Fisher information $I_{\theta_b}$ and the classical Fisher information $F^{\text{HOM}}_{\theta_b}$ are calculated as described in \secref{sec:metrology}, i.e., by taking the derivatives with respect to the estimated phase $\theta_b$. Throughout this analysis, $\theta_a$ acts as a tunable working-point parameter, whereas $\theta_b$ remains the phase to be estimated.

Figure~\ref{fig:HOM-metrology-thetaab} shows how the reference phase $\theta_a$ affects the sensitivity to $\theta_b$ and the information accessible through the binary HOM measurement. Figure~\figpanelNoPrefix{fig:HOM-metrology-thetaab}{a} shows the coincidence probability as a function of $\theta_b$ at fixed $\theta_a=0$, corresponding to the black dashed cut in the probability map in  \figpanel{fig:HOM-metrology-thetaab}{b}. The dependence of the coincidence probability on $\theta_a$ in \figpanel{fig:HOM-metrology-thetaab}{b} reflects how the reference phase modifies the GA self-interference and hence the effective two-photon beam-splitting process.

For the same reference phase $\theta_a=0$, \figpanel{fig:HOM-metrology-thetaab}{c} compares $I_{\theta_b}$ and $F^\text{HOM}_{\theta_b}$ as functions of $\theta_b$. The QFI characterizes the intrinsic sensitivity of the full two-photon output state to small variations of $\theta_b$, whereas the CFI quantifies the information accessible through the binary HOM measurement. 
Figure~\figpanelNoPrefix{fig:HOM-metrology-thetaab}{d} extends this comparison to different reference phases by showing the ratio $F^{\text{HOM}}_{\theta_b}/I_{\theta_b}$ as a function of $\theta_a$ and $\theta_b$. Values close to unity indicate that the binary HOM measurement  nearly saturates the QFI, i.e., that this measurement is nearly optimal for estimating $\theta_b$, whereas smaller values indicate that a larger fraction of the information encoded in the output state is not accessible through this measurement. A high ratio alone, however, does not imply high absolute sensitivity: favorable operating points also require a large CFI. The reference phase $\theta_a=\pi/2$, considered in the main text, provides one favorable choice, while other values between $0$ and $\pi/2$ can also be useful depending on the value of $\theta_b$ to be estimated.

\bibliography{ref}


\end{document}